\documentclass[twocolumn]{aastex631}
\usepackage{amsmath}
\usepackage{natbib}

\newcommand{\GG}[1]{}

\begin{document}

\title{ALMAGAL VII. Cataloging Hierarchical Continuum Structure from Cores to Clumps\\ across the Galactic Disk}

\shorttitle{ALMAGAL VII: Hierarchical Continuum Structure Catalog}
\shortauthors{Wallace et al.}

\correspondingauthor{Jennifer Wallace}
\email{jennifer.2.wallace@uconn.edu}

\author[0009-0002-7459-4174]{Jennifer Wallace}
\affiliation{University of Connecticut, Department of Physics, 196A Auditorium Road Unit 3046, Storrs, CT 06269 USA}

\author{Taevis Kolz}
\affiliation{University of Oregon, Department of Physics, 1371 E 13th Ave 120, Eugene, OR 97403 USA}
\affiliation{University of Connecticut, Department of Physics, 196A Auditorium Road Unit 3046, Storrs, CT 06269 USA}

\author[0000-0002-6073-9320]{Cara Battersby}
\affiliation{University of Connecticut, Department of Physics, 196A Auditorium Road Unit 3046, Storrs, CT 06269 USA}

\author[0000-0002-6946-6787]{Aleksandra Kuznetsova}
\affiliation{University of Connecticut, Department of Physics, 196A Auditorium Road Unit 3046, Storrs, CT 06269 USA}

\author[0000-0002-3078-9482]{\'Alvaro S\'anchez-Monge}
\affiliation{Institut de Ci\`encies de l'Espai (ICE), CSIC, Campus UAB, Carrer de Can Magrans s/n, E-08193, Bellaterra, Barcelona, Spain}
\affiliation{Institut d'Estudis Espacials de Catalunya (IEEC), E-08860, Castelldefels, Barcelona, Spain}

\author[0000-0003-1560-3958]{Eugenio Schisano}
\affiliation{INAF-Istituto di Astrofisica e Planetologia Spaziale, Via Fosso del Cavaliere 100, I-00133 Roma, Italy}

\author[0000-0001-8239-8304]{Alessandro Coletta}
\affiliation{INAF-Istituto di Astrofisica e Planetologia Spaziale, Via Fosso del Cavaliere 100, I-00133 Roma, Italy}
\affiliation{Dipartimento di Fisica, Sapienza Universit\`a di Roma, Piazzale Aldo Moro 2, I-00185, Rome, Italy}

\author[0000-0003-2384-6589]{Qizhou Zhang}
\affiliation{Center for Astrophysics $\vert$ Harvard \& Smithsonian, 60 Garden Street, Cambridge, MA, 02138, USA}

\author[0000-0002-9826-7525]{Sergio Molinari}
\affiliation{INAF-Istituto di Astrofisica e Planetologia Spaziale, Via Fosso del Cavaliere 100, I-00133 Roma, Italy}

\author[0000-0003-2141-5689]{Peter Schilke}
\affiliation{I. Physikalisches Institut, Universit{\"a}t zu K{\"o}ln, Z{\"u}lpicher Stra{\ss}e 77, 50937 Cologne, Germany}

\author[0000-0002-3412-4306]{Paul T. P. Ho}
\affiliation{Institute of Astronomy and Astrophysics, Academia Sinica, 11F of ASMAB, AS/NTU No.\ 1, Sec.\ 4, Roosevelt Road, Taipei 10617, Taiwan}
\affiliation{East Asian Observatory, 660 N.\ A'ohoku, Hilo, Hawaii, HI 96720, USA}

\author[0000-0003-2309-8963]{Rolf Kuiper}
\affiliation{Faculty of Physics, University of Duisburg-Essen, Lotharstra{\ss}e 1, D-47057 Duisburg, Germany}

\author[0000-0002-1466-3484]{Tianwei Zhang}
\affiliation{Research Center for Earth and Space Computing, Zhejiang Laboratory, Hangzhou, China}
\affiliation{I. Physikalisches Institut, Universit{\"a}t zu K{\"o}ln, Z{\"u}lpicher Stra{\ss}e 77, 50937 Cologne, Germany}

\author[0000-0002-9277-8025]{Thomas M\"{o}ller}
\affiliation{I. Physikalisches Institut, Universit{\"a}t zu K{\"o}ln, Z{\"u}lpicher Stra{\ss}e 77, 50937 Cologne, Germany}

\author[0000-0002-0560-3172]{Ralf S.\ Klessen}
\affiliation{Universit\"{a}t Heidelberg, Zentrum f\"{u}r Astronomie, Institut f\"{u}r Theoretische Astrophysik, Albert-Ueberle-Str.\ 2, 69120 Heidelberg, Germany}
\affiliation{Universit\"{a}t Heidelberg, Interdisziplin\"{a}res Zentrum f\"{u}r Wissenschaftliches Rechnen, Im Neuenheimer Feld 225, 69120 Heidelberg, Germany}

\author[0000-0003-3315-5626]{Maria T.\ Beltr\'an}
\affiliation{INAF-Osservatorio Astrofisico di Arcetri, Largo E.\ Fermi 5, I-50125 Firenze, Italy}

\author[0000-0002-8942-1594]{Floris van der Tak}
\affiliation{Space Research Organization Netherlands (SRON), Landleven 12, 9747 AD Groningen, NL }
\affiliation{Kapteyn Astronomical Institute, University of Groningen, NL}

\author[0000-0001-7852-1971]{Stefania Pezzuto}
\affiliation{INAF-Istituto di Astrofisica e Planetologia Spaziale, Via Fosso del Cavaliere 100, I-00133 Roma, Italy}

\author[0000-0002-1700-090X]{Henrik Beuther}
\affiliation{Max Planck Institute for Astronomy, Koenigstuhl 17, 69117 Heidelberg, Germany}

\author[0000-0003-1665-6402]{Alessio Traficante}
\affiliation{INAF-Istituto di Astrofisica e Planetologia Spaziale, Via Fosso del Cavaliere 100, I-00133 Roma, Italy}

\author[0000-0002-9120-5890]{Davide Elia}
\affiliation{INAF-Istituto di Astrofisica e Planetologia Spaziale, Via Fosso del Cavaliere 100, I-00133 Roma, Italy}

\author[0000-0002-9574-8454]{Leonardo Bronfman}
\affiliation{Departamento de Astronomía, Universidad de Chile, Casilla 36-D, Santiago, Chile}

\author[0000-0001-9443-0463]{Pamela Klaassen}
\affiliation{UK Astronomy Technology Centre, Royal Observatory Edinburgh, Blackford Hill, Edinburgh EH9 3HJ, UK}

\author[0000-0002-0500-4700]{Dariusz C. Lis}
\affiliation{Jet Propulsion Laboratory, California Institute of Technology, 4800 Oak Grove Drive, Pasadena, CA 91109, USA}

\author[0000-0002-8517-8881]{Luca Moscadelli}
\affiliation{INAF-Osservatorio Astrofisico di Arcetri, Largo E.\ Fermi 5, I-50125 Firenze, Italy}

\author[0000-0003-4146-9043]{Kazi Rygl}
\affiliation{INAF-Istituto di Radioastronomia \& Italian ALMA Regional Centre, Via P. Gobetti 101, I-40129 Bologna, Italy}

\author[0000-0002-3597-7263]{Milena Benedettini}
\affiliation{INAF-Istituto di Astrofisica e Planetologia Spaziale, Via Fosso del Cavaliere 100, I-00133 Roma, Italy}

\author[0000-0003-1964-970X]{Chi Yan Law}
\affiliation{INAF-Osservatorio Astrofisico di Arcetri, Largo E.\ Fermi 5, I-50125 Firenze, Italy}

\author[0009-0007-4060-0560]{Jofre Allande}
\affiliation{INAF-Osservatorio Astrofisico di Arcetri, Largo E.\ Fermi 5, I-50125 Firenze, Italy}
\affiliation{Dipartimento di Fisica e Astronomia, Università degli Studi di Firenze, Via G. Sansone 1, 50019 Sesto Fiorentino, Firenze, Italy}

\author[0009-0005-9192-5491]{Alice Nucara}
\affiliation{INAF-Istituto di Astrofisica e Planetologia Spaziale, Via Fosso del Cavaliere 100, I-00133 Roma, Italy}
\affiliation{Dipartimento di Fisica, Universit\`a di Roma Tor Vergata, Via della Ricerca Scientifica 1, I-00133 Roma, Italy}

\author[0000-0003-2777-5861]{Patrick M. Koch}
\affiliation{Institute of Astronomy and Astrophysics, Academia Sinica, 11F of ASMAB, AS/NTU No.\ 1, Sec.\ 4, Roosevelt Road, Taipei 10617, Taiwan}

\author[0000-0003-0364-6715]{Won-ju Kim}
\affiliation{I. Physikalisches Institut, Universit{\"a}t zu K{\"o}ln, Z{\"u}lpicher Stra{\ss}e 77, 50937 Cologne, Germany}
\affiliation{Max-Planck-Institut f\"{u}r Radioastronomie, Auf dem H\"{u}gel 69, 53121 Bonn, Germany}

\author[0000-0002-7125-7685]{Patricio Sanhueza}
\affiliation{Department of Astronomy, School of Science, The University of Tokyo, 7-3-1 Hongo, Bunkyo, Tokyo 113-0033, Japan}

\author[0000-0001-8509-1818]{Gary Fuller}
\affiliation{Jodrell Bank Centre for Astrophysics, Oxford Road, The University of Manchester, Manchester M13 9PL, UK}
\affiliation{I. Physikalisches Institut, Universit{\"a}t zu K{\"o}ln, Z{\"u}lpicher Stra{\ss}e 77, 50937 Cologne, Germany}

\author[0000-0002-4935-2416]{Georgie Stroud}
\affiliation{Jodrell Bank Centre for Astrophysics, Oxford Road, The University of Manchester, Manchester M13 9PL, UK}

\author[0000-0002-0675-0078]{Beth Jones}
\affiliation{I. Physikalisches Institut, Universit{\"a}t zu K{\"o}ln, Z{\"u}lpicher Stra{\ss}e 77, 50937 Cologne, Germany}

\author[0000-0002-6558-7653]{Crystal Brogan}
\affiliation{National Radio Astronomy Observatory, 520 Edgemont Road, Charlottesville VA 22903, USA}

\author[0000-0001-6492-0090]{Todd Hunter}
\affiliation{National Radio Astronomy Observatory, 520 Edgemont Road, Charlottesville VA 22903, USA}

\author[0000-0003-4037-5248]{Aida Ahmadi}
\affiliation{Leiden Observatory, Leiden University, PO Box 9513, 2300 RA Leiden, The Netherlands}

\author[0000-0002-2562-8609]{Adam Avison}
\affiliation{SKA Observatory, Jodrell Bank, Lower Withington, Macclesfield, SK11 9FT, UK}
\affiliation{Jodrell Bank Centre for Astrophysics, Oxford Road, The University of Manchester, Manchester M13 9PL, UK}
\affiliation{UK ALMA Regional Centre Node, M13 9PL, UK}

\author[0000-0003-4509-1180]{Katharine Johnston}
\affiliation{School of Engineering and Physical Sciences, Isaac Newton Building, University of Lincoln, Brayford Pool, Lincoln, LN6 7TS, United Kingdom}

\author{Sheng-Yuan Liu}
\affiliation{Institute of Astronomy and Astrophysics, Academia Sinica, 11F of ASMAB, AS/NTU No.\ 1, Sec.\ 4, Roosevelt Road, Taipei 10617, Taiwan}

\author[0000-0002-2974-4703]{Chiara Mininni}
\affiliation{INAF-Istituto di Astrofisica e Planetologia Spaziale, Via Fosso del Cavaliere 100, I-00133 Roma, Italy}

\author{Yu-Nung Su}
\affiliation{Institute of Astronomy and Astrophysics, Academia Sinica, 11F of ASMAB, AS/NTU No.\ 1, Sec.\ 4, Roosevelt Road, Taipei 10617, Taiwan}

\author{Hans Zinnecker}
\affiliation{Universidad Autonoma de Chile, Avda Pedro de Valdivia 425, Santiago de Chile}

\begin{abstract}
Investigating the multi-scale fragmentation of dense clumps into compact cores is essential for understanding the processes that govern the initial distribution of mass in stellar clusters and how high-mass stars ($>8~M_{\odot}$) form. We present a catalog of the hierarchical continuum structure from 904 clumps observed in the ALMAGAL program, a high resolution ($0.15-0.8$\arcsec) 1.38 mm Atacama Large Millimeter/submillimeter Array (ALMA) large program targeting dense clumps capable of high-mass star formation throughout the Galactic disk. We use \verb|astrodendro|, a dendrogram-based algorithm, on a uniform linear resolution (2000 au) version of the data to extract 5160 continuum structures with effective radii spanning $800-42000$ au and estimated masses between $~0.05-670~M_{\odot}$. With our large sample, we statistically examine differences in clump properties for regions with varying levels of hierarchical complexity. We find that clumps exhibiting the richest hierarchical morphology have distributions with higher dust temperatures, surface densities, luminosity-to-mass (\textit{L/M}) ratios, and most massive core (MMC) masses, indicating that these regions tend to be at later evolutionary stages. We find a positive correlation between the mass of cores from the ALMAGAL core catalog and the surface density of their surrounding structures identified in this work. However, this correlation is weaker for cores in more evolved clumps, where lower mass cores can be found at higher local surface densities. This could indicate that some cores accrete mass less efficiently from the intra-clump reservoir than others, despite the total available mass increasing over time, a scenario that is congruent with a clump-fed core accretion model.
\end{abstract}


\section{Introduction}
\label{intro}

Star formation occurs within a hierarchically structured interstellar medium composed of giant molecular clouds ($\sim$10--100 pc) that fragment into dense clumps ($\sim$ 1 pc) and then cores ($\sim$ 0.01--0.1 pc). These dense cores are the progenitors of  individual stars or groups of stars \citep{lada_lada_2003}. Approximately 50\% of these stars form in densely populated ($>$1000 stars) clusters containing at least one 10 M$_{\odot}$ star \citep{Adams_2010}. The stellar feedback from these high-mass stars is expected to have a large impact on the formation of stars in their immediate environment via gravitational interactions as well as mechanical and radiative feedback \citep[e.g.][]{Rosen_Krumholz_2020}.

There are competing theories on whether high-mass stars form directly from a massive `seed' core, similar to their isolated, lower mass counterparts in a \textit{core-fed} scenario \citep[e.g.]{McKee_Tan_2003, Tan_McKee_2003, Krumholz_2005} or whether they dynamically accrete material from a larger scale mass reservoir in a \textit{clump-fed} scenario \citep[e.g.][]{Zinnecker_1982, Bonnell_2001, Bonnell_Bate_2006, Girichidis_2011, Vasquez_Semadeni_2019, Padoan_2020, Anderson_2021, Traficante_2023}. These dynamical accretion models differ in their expression. For example, in a competitive accretion model, cores that initially accrete mass from the intra-clump medium will continue to do so efficiently as their gravitational influence becomes more significant within the clump \citep{Bonnell_2001, Bonnell_Bate_2006}. However, in the fragmentation-induced starvation model, cores at the global minimum of the gravitational potential in a clump will accrete mass less efficiently as outer cores prevent some intra-clump gas from reaching the innermost cores \citep{Thomas_2010, Girichidis_2012}.

There are many high resolution star formation studies that have been performed locally, in environments where stars form in isolation or within small clusters \citep[e.g.][]{Enoch_2007, Andre_2010, 2023_Gomez, 2025_Cesaroni}. However, due to observational limitations, it is more difficult to investigate star-forming regions with heliocentric distances $>500$ pc at a similar resolution. 

Through the use of state-of-the-art facilities such as ALMA, it has become possible to probe regions capable of forming high-mass stars at the resolution and sensitivity required to study the process of star formation (1000s au). Some studies have focused on regions in the early phases of star formation, such as Submillimeter Array (SMA) and ALMA observations of Infrared Dark Clouds (IRDCs) \citep{Zhang_2009,Zhang_2015}, and the ALMA Survey of 70\micron ~Dark High-mass Clumps in Early Stages \citep[ASHES,][]{Sanhueza_2019}. Others have instead investigated more evolved star-forming clumps, such as the CORE \citep{Beuther_2018}, ATOMS \citep{Liu_2020} and ASSEMBLE surveys \citep{Xu_2024}. Several surveys that have probed star formation across a range of evolutionary stages include SQUALO \citep{Traficante_2023}, TEMPO \citep{Avison_2023}, and DIHCA \citep{Ishihara_2024}.

However, these surveys have been limited in statistical relevance, typically analyzing only tens of star-forming regions at a time. Additionally, these studies have been conducted with observations at different sensitivities and spatial resolutions, and have extracted compact sources using different algorithms, such as \textit{CuTEx} \citep{Molinari_2011}, \textit{astrodendro} \citep{Robitaille_2019}, \textit{getsf} \citep{Menshchikov_2021}, and \textit{hyper} \citep{Traficante_2015}, which all have different assumptions regarding the morphology of the source being extracted. This lack of observational and methodological consistency between surveys makes it difficult to draw conclusions when comparing scientific results.  

The ALMA evolutionary study of high-mass protocluster formation in the GALaxy (ALMAGAL) survey \citep{Molinari_2025, Sanchez-Monge_2025} allows us to investigate star formation across the Galactic disk using a large and statistically representative sample of high-mass star-forming regions that can be uniformly analyzed. This large ALMA program includes 1.38mm continuum and spectral line observations towards more than 1000 high-mass and dense Galactic clumps. These regions have been specifically selected to cover a wide range of Galactic environments and span the full range of clump evolution, from IRDCs to H{\sc ii} regions. Initial results from the first round of ALMAGAL papers have already given us significant insight on the properties of cores throughout the Galactic disk, and how their fundamental properties and spatial distribution relate to the evolutionary stage of their parental clumps and immediate clump environment \citep[e.g.][]{Coletta_2025, Mininni_2025, Schisano_2025, Elia_2026}. Additionally, the ALMAGAL survey has synergy with other existing high-resolution and statistically robust data sets, such as the ALMA-IMF survey \citep{Motte_2022}, which leverages different observing strategies by covering a large area around 15 massive star-forming clumps to investigate the origin of stellar clusters and the initial mass function. 

In this paper, we catalog the hierarchy of continuum structures observed in the ALMAGAL survey and investigate the multi-scale fragmentation of clumps across a range of evolutionary stages. It is organized as follows: in Section \ref{sec:obs} we describe the  ALMAGAL survey observations and the methods used to acquire our final data products. In Section \ref{sec:methods} we outline the procedure used to catalog multi-scale continuum structures and the methods used for calculating their properties. We also define our method for classifying regions based on their relative level of hierarchical fragmentation. In Section \ref{sec:results} we report on the resulting continuum structure properties and show how the relative level of hierarchical structure seen in the ALMAGAL targets is correlated with its clump-scale properties. In Section \ref{sec:discussion} we discuss our results in the context of Galactic star formation, with a focus on how core mass accretion proceeds at different evolutionary stages. Finally, Section \ref{sec:conclusion} is a brief summary of our conclusions.

\begin{figure*}[t!] 
\epsscale{1.1}
\begin{centering}
\plotone{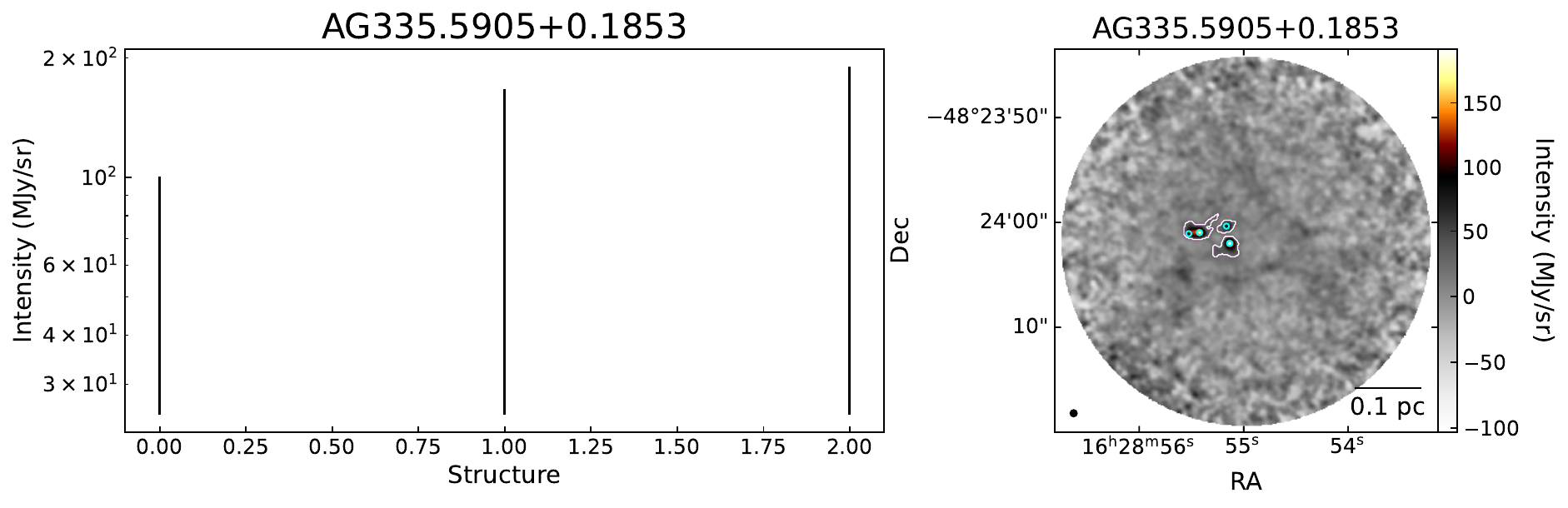}
\plotone{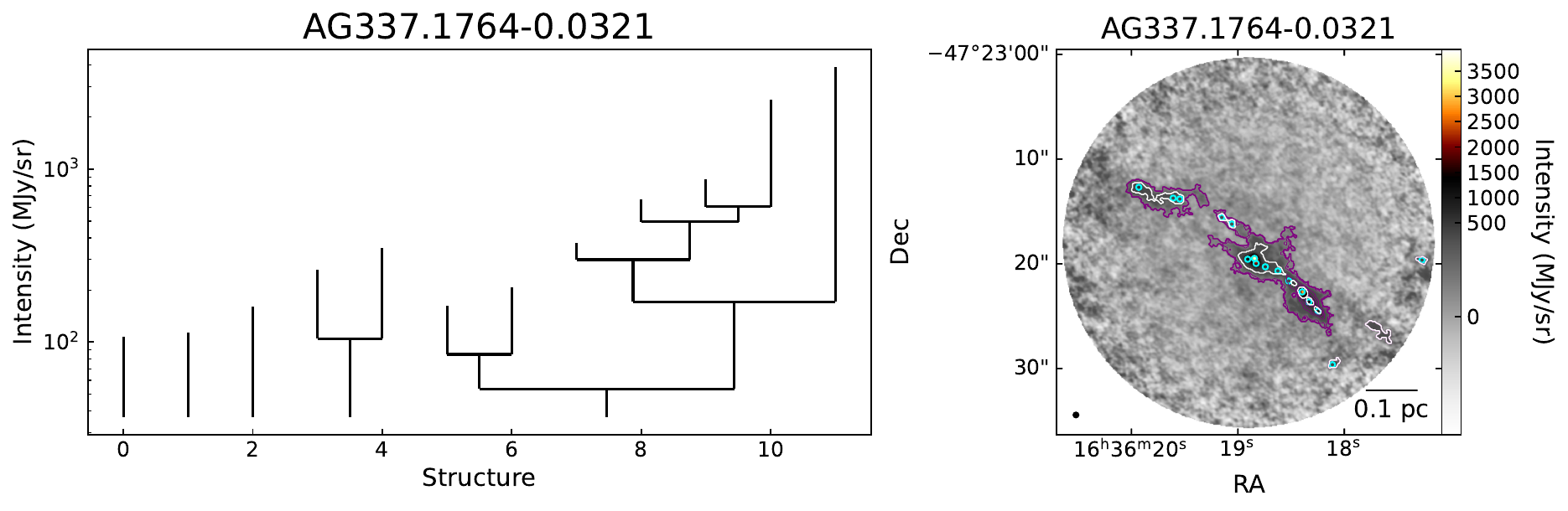}
\plotone{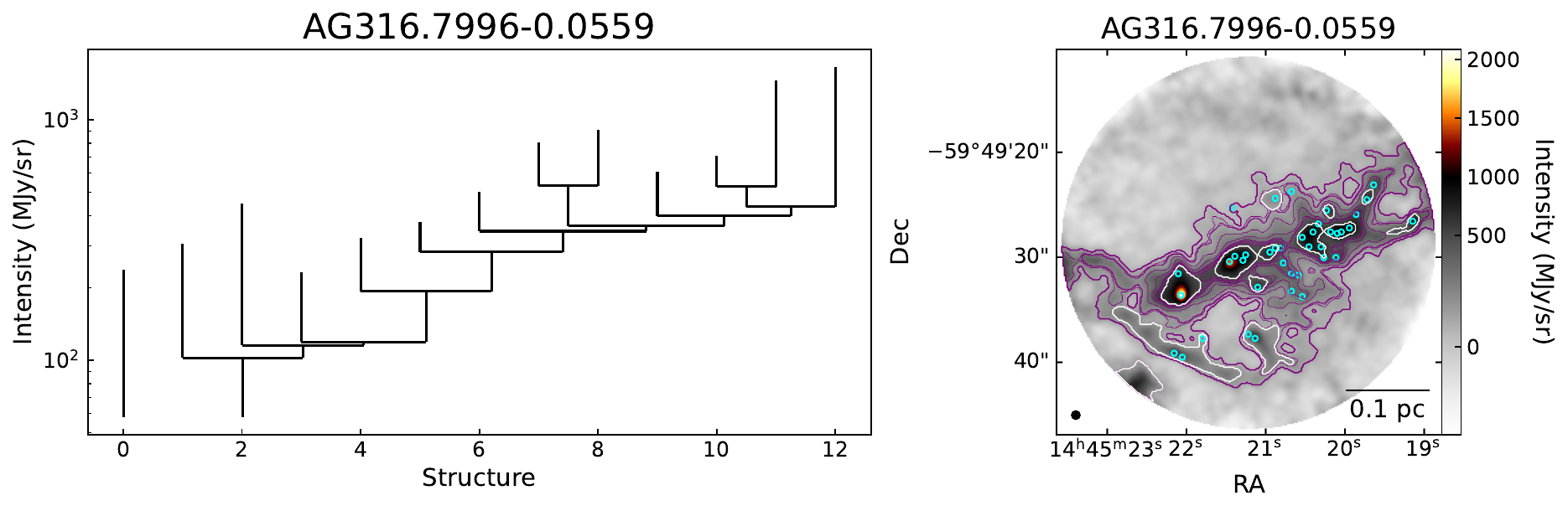}
\caption{Three ALMAGAL targets are selected as examples to depict typical morphology in the Isolated (top), Simple (middle), and Rich (bottom) clump categories. The method used to classify these targets is described in Section \ref{subsec:clump_class}. The ID for each target is given at the top of each panel, and it corresponds to the ``ALMAGAL ID'' panel in Table \ref{tab:catalog}. In the right panels, background grayscale fading into heat colorbar shows the ALMAGAL 1.3mm dust continuum emission. Purple contours indicate the branch structures identified by the dendrogram algorithm in each region, while white contours indicate the leaf structures. Cyan circles indicate the position of cores from the ALMAGAL core catalog \citep{Coletta_2025}. The left panel in each figure shows the dendrogram structure produced from the images shown in the corresponding right panels using the methods described in Section \ref{subsec:dendro}. The size of the synthesized 2000 AU beam is shown in the bottom left corner of each image and a scale bar is included in the bottom right corner. }
\label{fig:clump_class}
\end{centering}
\end{figure*}

\section{Data}
\label{sec:obs}
In this section we provide a brief description of the ALMAGAL survey target sample, the properties of the continuum observations, our procedure for producing a uniform sub-sample from the data set, and ancillary data sets used in our analysis. For a more complete description of the ALMAGAL target sample as well as the properties of the original continuum data products, we defer to the ALMAGAL Survey overview paper \cite{Molinari_2025} and data overview paper \cite{Sanchez-Monge_2025}.

\subsection{The ALMAGAL Survey Target Sample}
\label{subsec:alma_samp}

The ALMAGAL target sample consists of 1013 dense clumps located throughout the Galactic disk. These regions are candidates for hosting high-mass star formation and were selected from 915 clumps identified in the Herschel Infrared Galactic Plane Survey \citep[Hi-GAL,][]{Molinari_2010, Molinari_2016b, Elia_2017, Elia_2021}, and 98 clumps from the Red MSX Survey \citep[RMS,][]{Lumsden_2013}. The initial proposed target selection criteria required that clumps have a mass $>$500 $M_{\odot}$ for targets in the inner Galaxy and $>$250 $M_{\odot}$ for those in the outer Galaxy. Observed targets were also required to have a surface density $\Sigma \geq$ 0.1 g cm$^{-2}$, a threshold designed to include systems where high-mass stars are likely to form \citep{Krumholz_McKee_2008}. 
    
After a revision of the heliocentric distance estimates made after the ALMAGAL targets were observed, there are some targets that no longer fulfill these selection criteria. However, 90\% of the targets still meet these initial selection requirements \citep{Molinari_2025}. Target clumps span a wide range of heliocentric distances (0.5-10 kpc) and Galactocentric radii (3-14 kpc).

In Section \ref{subsec:clump_class}, we evaluate the results from the continuum structure extraction as it relates to the properties of their host clumps, as they are reported in the revised and consolidated version of the ALMAGAL property table in \cite{Molinari_2025}\footnote{This table is available at the CDS via the following link: \url{https://
cdsarc.cds.unistra.fr/viz-bin/cat/J/A+A/696/A149}}. We note that there are 6 clumps for which an SED fit could not be used to make a dust temperature estimate, so they do not have reported masses or surface densities.

\subsection{ALMAGAL Survey Observations}
\label{subsec:alma_obs}
In this paper, we use the fully combined (7M+TM2+TM1)\footnote{This label refers to the fully combined data products obtained using the 7M (7m-array), TM2 (C-2 \& C-3) and TM1 (C-5 \& C-6) ALMA array configurations.} ALMAGAL 1.38 mm continuum data products \citep{Sanchez-Monge_2025}. The full set of targets is divided by their heliocentric distances into a `near' ($d \leq 4.7$ kpc) and `far' sample ($d > 4.7$ kpc). These two sub-samples were observed with different antenna configurations to achieve a more uniform linear resolution for all targets. The median angular resolution for the near and far sample ALMAGAL images are $0.47\arcsec \times 0.38\arcsec$ and $0.28\arcsec \times 0.19\arcsec$, respectively. The largest angular scale (LAS) for all of the ALMAGAL observations is $\sim$ 29\arcsec. The rms noise values for each of the continuum maps ($\sigma_{\text{rms}}$) were estimated by taking the standard deviation of the residual images after they were masked to exclude regions of bright emission \citep[see][]{Sanchez-Monge_2025}.

\subsection{Uniform Resolution ALMAGAL Continuum Maps}
\label{subsec:smooth}

To ensure that our results are consistent for the entire ALMAGAL sample, we created a uniform linear resolution subset of the continuum images. First, we converted the native resolution images to the beam independent flux density unit of MJy/sr. Then, we calculated the circularized linear beam size for each image by taking the geometric mean of the major and minor axes of the beam. We then removed all targets with a corresponding linear spatial resolution $> 2000$ au, and smoothed the remaining 904 images with a 2D Gaussian kernel so that each source had a circularized beam with a linear size of 2000 au. We chose the 2000 au threshold to keep the majority of the full ALMAGAL sample ($\sim90\%$), without reducing the spatial resolution of the data by a substantial amount. Additionally, we downsampled each smoothed image so that the 2000 au beam was sampled by 6 pixels.  We note that only 900 of these targets have associated measured clump properties and consequently do not use the remaining 4 in the analysis presented in Section \ref{subsec:clump_distributions}. The uniform linear
resolution version of the ALMAGAL continuum images are
available at Zenodo doi:\url{10.5281/zenodo.18421180}. 

\begin{table*}
\caption{The first 10 rows of the ALMAGAL hierarchical structure catalog. The full catalog is available at Zenodo doi:\url{10.5281/zenodo.18421180}.} 
\begin{center}
\begin{tabular}{cccccccccccc}
\hline\hline
ID & ALMAGAL ID & RA & Dec & R$_{\text{eff}}$ & $F_{\nu}$ & $F_{\nu \text{,bgs}}$ & $I_{\text{peak}}$ & $\sigma_{\text{rms}}$ & Structure & Level & Type \\
 &  & (deg) & (deg) & (au) & (mJy) & (mJy) & (MJy sr$^{-1}$) & (MJy sr$^{-1}$) & &  &\\
\hline
0 & AG022.7068+0.3897 & 277.806 & -8.909 & 2349 & 2.56 & 2.56 & 5.74e+02 & 39.79 & Leaf & 0 & isolated \\
1 & AG022.7068+0.3897 & 277.807 & -8.909 & 2103 & 1.31 & 1.31 & 2.67e+02 & 39.79 & Leaf & 0 & isolated \\
2 & AG022.7068+0.3897 & 277.807 & -8.908 & 3827 & 4.18 & 4.18 & 2.47e+02 & 39.79 & Leaf & 0 & isolated \\
3 & AG022.7384-0.2420 & 278.389 & -9.176 & 1823 & 1.99 & 1.99 & 3.67e+02 & 50.03 & Leaf & 0 & simple \\
4 & AG022.7384-0.2420 & 278.390 & -9.174 & 10422 & 69.76 & 69.76 & 1.35e+03 & 50.03 & Branch & 0 & simple \\
5 & AG022.7384-0.2420 & 278.390 & -9.174 & 3300 & 15.85 & 9.07 & 1.35e+03 & 50.03 & Leaf & 2 & simple \\
6 & AG022.7384-0.2420 & 278.390 & -9.174 & 5489 & 38.82 & 20.63 & 1.35e+03 & 50.03 & Branch & 1 & simple \\
7 & AG022.7384-0.2420 & 278.389 & -9.174 & 4264 & 22.31 & 11.02 & 1.18e+03 & 50.03 & Leaf & 2 & simple \\
8 & AG022.7384-0.2420 & 278.390 & -9.174 & 2272 & 4.31 & 1.09 & 4.71e+02 & 50.03 & Leaf & 1 & simple \\
9 & AG022.8624+0.4213 & 277.852 & -8.757 & 5154 & 10.63 & 10.63 & 7.09e+02 & 48.70 & Branch & 0 & simple \\
\hline
\end{tabular}
\end{center}
\tablecomments{For each dendrogram structure we report the structure ID, ALMAGAL clump ID, central coordinates in degrees (RA,Dec), effective radius in au ($R_{\text{eff}}$), integrated flux in Jy ($F_{\nu}$), background subtracted integrated flux in Jy ($F_{\nu \text{,bgs}}$), peak intensity in MJy sr$^{-1}$ ($I_{\text{peak}}$), the RMS noise of the continuum map in MJy sr$^{-1}$ ($\sigma_{\text{rms}}$), the structure type (leaf or branch), its dendrogram level, and the clump type, which is its morphological classification (Empty, Isolated, Simple, Rich).}
\label{tab:catalog}

\end{table*}

\subsection{The ALMAGAL Compact Core Catalog}
\label{subsec:core_cat}

As the hierarchical structure catalog we present in this paper is not designed to identify individual cores, we instead use the ALMAGAL compact core catalog \citep{Coletta_2025}  for information on the position, sizes, and masses of cores found in the ALMAGAL sample. This information is useful for understanding the placement of cores within the continuum structure hierarchy in each image. For this catalog, the core extraction was performed using a version of the CuTEx algorithm \citep{Molinari_2011},  that was modified to be used with the ALMAGAL continuum maps. We note that the 1013 targets from the original (non-smoothed) continuum data products were used to generate this catalog. A detailed description of the source detection procedure and calculations of core properties can be found in \cite{Coletta_2025}. 

In total there are 6348 sources included in the ALMAGAL core catalog, with 5487 sources contained in our uniform linear resolution sample of 900 clumps. The physical size distribution is between 200 - 10000 au, with 90\% of cores ranging from $\sim 800-3000$ au, with a median value of 1700 au. The calculated core masses span $0.002 - 345 ~M_{\odot}$, with a median value of 0.4 $M_{\odot}$. As described in \cite{Coletta_2025}, masses were calculated using the modified blackbody formula, assuming a dust opacity of 0.9 g cm$^{-2}$ and a gas to dust ratio of 100. The core temperatures were estimated according to the $L/M$ ratio of its host clump (see \cite{Coletta_2025} for details). We utilize this same prescription for calculating the mass of our dendrogram structures, as described in Section \ref{subsec:struc_prop}. 

In this paper, we report the MMC mass from each clump using the core masses from the ALMAGAL compact core catalog. Additionally, we note that we use the terminology of ``core'' to specifically refer to the cores from the ALMAGAL core catalog, and not objects from the hierarchical structure catalog presented here.

\subsection{Ancillary Data: ATLASGAL}
\label{subsec:atlasgal}
In Appendix \ref{sec:feathering}, we assess how feathering the ALMAGAL data with single-dish observations affects our results. For this we use data from the APEX Telescope Large Area Survey of the Galaxy (ATLASGAL), a Galactic plane survey spanning Galactic longitudes \textit{l}$=\pm60\degr$ and Galactic latitudes \textit{b}$=\pm1.5\degr$ \citep{Schuller_2009}. This survey was later extended to include $-80\degr<$ \textit{l} $<-60\degr$ and $-2\degr<$ \textit{b} $<1\degr$. The ATLASGAL dust continuum observations are made at 870 micron ($\sim$ 345 GHz). The APEX telescope has a full width at half-maximum (FWHM) beam size of 19\arcsec.2 at this wavelength. The data have a field of view (FOV) of 11$\arcmin$.4 and a typical noise level of 50-70 mJy/beam.

\begin{figure}[htb!]
\epsscale{1.0}
\plotone{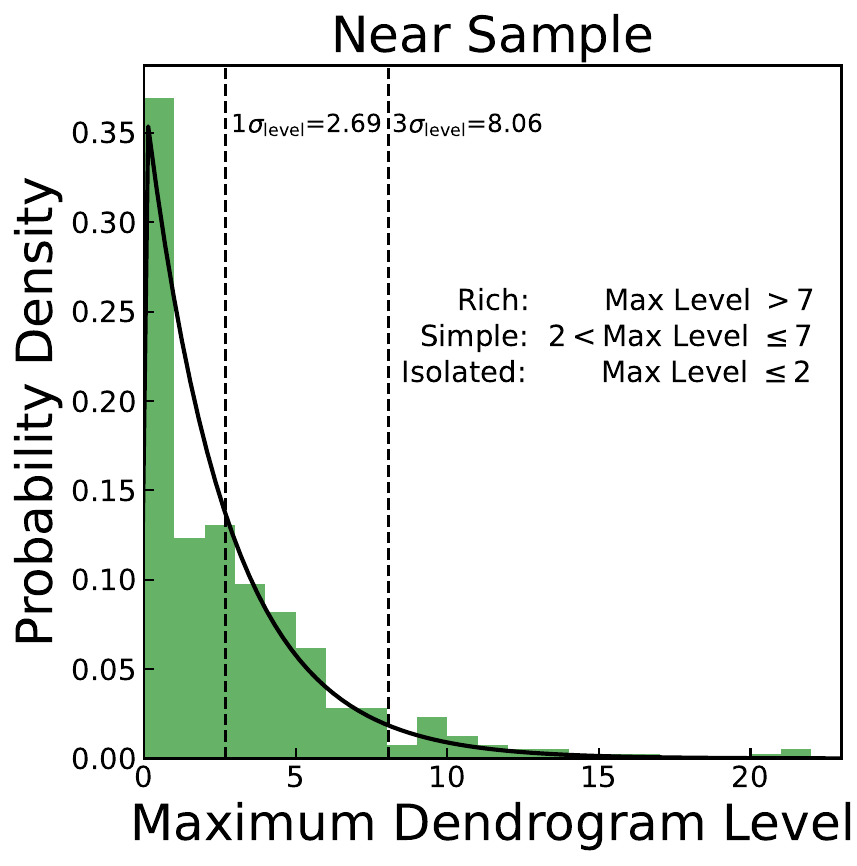}
\plotone{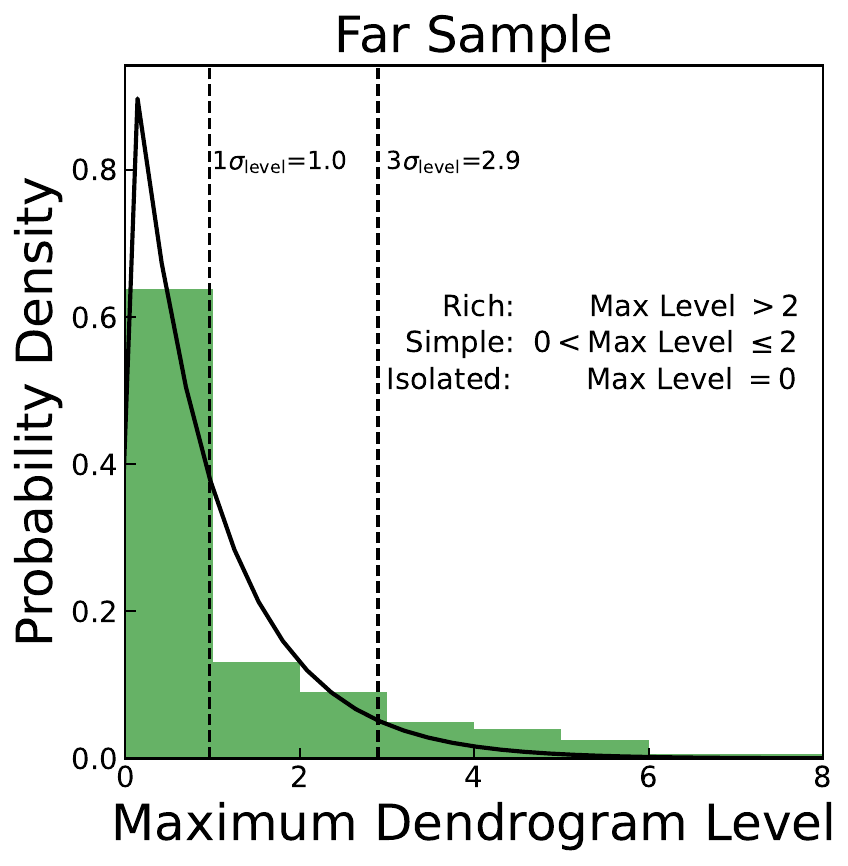}
\caption{The maximum dendrogram level distribution for the near (top panel) and far (bottom panel) clumps in our ALMAGAL sample. The vertical dashed lines indicate the 1$\sigma_{\text{level}}$ and 3$\sigma_{\text{level}}$ of the fitted exponential distribution, shown by the solid black line. Since we cannot use the exact 1$\sigma_{\text{level}}$ and 3$\sigma_{\text{level}}$ values as thresholds on the integer-valued dendrogram levels, we instead round each value to the nearest integer and use that as a threshold. The integer ranges for the maximum level are shown for the Rich, Simple, and Isolated targets. The Empty targets do not have a threshold value, since they represent regions with non-detections and do not have an associated dendrogram level.}
\label{fig:max_lvl_dists}
\end{figure}

\section{Methods}
\label{sec:methods}

\subsection{Dendrogram Algorithm}
\label{subsec:dendro}
To generate our hierarchical structure catalog, we implement a dendrogram-based algorithm using the Python package \verb|astrodendro| \footnote{Detailed information on the \texttt{astrodendro} Python package can be found at \url{http://www.dendrograms.org/}.}. Dendrograms are tree-like diagrams used to represent the hierarchical structure of multi-level data \citep{Rosolowsky_2008}. The densest structures that contain no sub-structure are called `leaves'. The structures containing sub-structures are referred to as `branches' and the structures that have no parent structures are referred to as `trunks'. Each structure has a `level' associated with it. All trunks have a level = 0, the first level of sub-structure after the trunk has a level = 1, and it increases until a maximum level is reached at a terminating leaf structure.

Our decision to use a dendrogram-based approach is because it is uniquely suited to identifying hierarchical structures without assuming an underlying morphology, such as a compact 2D Gaussian or an elongated filament. This being the case, we emphasize that the leaf structures presented in our catalog are \textit{not} directly comparable to the pre- or protostellar core structures identified in the ALMAGAL compact core catalog \citep{Coletta_2025}.

We run the \verb|astrodendro| algorithm on images that have not undergone primary beam correction, since the noise fluctuations are homogenous across the maps. However, the flux measurements for these structures are calculated using the primary beam corrected images. 

We generate a dendrogram for each image using four parameters: 

\begin{enumerate}
    \item Minimum structure value, $f_{\text{min}}$: the lowest allowed peak pixel value for a given structure.
    \item Minimum significance, $\delta$: the threshold for how much brighter than the surrounding structure a leaf must be in order to be considered an independent structure
    \item Minimum peak value, $f_{\text{peak}}$: the minimum allowed value that the peak intensity of a structure can have.
    \item Minimum number of pixels, $n_{\text{pix}}$: the minimum number of pixels required for a structure to be included in the dendrogram. 
\end{enumerate}

We scale our inputs for these parameters using the rms noise ($\sigma_{\text{rms}}$) for each image so that $f_{\text{min}}=2\sigma_{\text{rms}}$, $f_{\text{peak}}=3\sigma_{\text{rms}}$, and $\delta=3\sigma_{\text{rms}}$. The $\sigma_{\text{rms}}$ noise values correspond to the `AGSTDREM' values reported for the original, non-smoothed images from \cite{Sanchez-Monge_2025}, converted into units of MJy/sr. We take $n_{\text{pix}}$ to be equal to $0.5 \times A_{\text{beam,pix}}$, where $A_{\text{beam,pix}}$ is the number of pixels contained within the Gaussian area of the synthesized beam for each image. We choose a factor of 0.5 instead of 1.0 because the area contained within the structure boundary at or above $f_{\text{min}}=2\sigma_{\text{rms}}$ may not exceed the full area of the beam, even if it is an otherwise significant detection. This can result in extracting structures that are slightly smaller than the beam, however this choice in  $n_{\text{pix}}$ allows us to recover significant detections that would otherwise be removed. The first 10 rows of the catalog are shown in Table \ref{tab:catalog}. In total we report 5160 structures in our catalog, 3470 of the structures are leaves and 1690 are branches. We present the dendrograms and corresponding structures for three different targets in Figure \ref{fig:clump_class}, where leaf and branch structures are shown in white and purple contours, respectively. The
output dendrogram files are available in FITS format on
Zenodo doi:\url{10.5281/zenodo.18421180}. The Jupyter notebooks used to generate the dendrogram as well as perform
subsequent analysis on the resulting catalog are available at
\url{https://github.com/Milky-Way-Laboratory/ALMAGAL}.

\begin{table}
\begin{center}
\caption{A cross-sectional summary of the 900 ALMAGAL targets with measured clump properties that we analyze in this paper. We consider the morphological sub-samples (Rich, Simple, Isolated, Empty) of clumps in the context of their \textit{L/M} ratios (in units of $L_{\odot}/M_{\odot}$), as well as whether they are included in the near or far sample observations.} 
\begin{tabular}{lcccc}
\hline\hline
Sample & $\textit{L/M} \leq 1$ &$1 < \textit{L/M} \leq 10$ & $\textit{L/M} > 10$ & All\\
\hline
\textbf{Rich} & 4 & 17 & 56 & 77\\
\hline
Near & 2 & 11 & 19 & 32\\
Far  & 2 & 6 & 37 &  45\\
\hline
\textbf{Simple}  & 37 & 82 & 68 & 187 \\
\hline 
Near  & 23 & 47 & 46 & 116 \\
Far  & 14 & 35 & 22 & 71 \\
\hline
\textbf{Isolated}  & 232 & 120 & 93 & 445\\
\hline
Near  & 141 & 47 & 54 & 242 \\
Far  & 91 & 73 & 39 & 203 \\
\hline
\textbf{Empty} & 118 & 61 &  12 & 191\\
\hline
Near & 47  & 7 & 0 & 54 \\
Far  & 71 & 54 & 12 & 137 \\
\hline
\textbf{All} & 391 & 280 & 229 & 900 \\
\hline
Near & 213 & 112 & 119 & 444 \\
Far &  178 & 168 & 110 & 456 \\
\hline
\end{tabular}
\label{tab:sample}
\end{center}
\end{table}

\subsection{Hierarchical Classification of the ALMAGAL Sample}
\label{subsec:clump_class}

We first consider the morphological complexity as a bulk property of each ALMAGAL target, or clump. As described in Section \ref{subsec:dendro}, each structure has a dendrogram `level' that provides a relative description of how deeply nested the structure is in relation to its parent structure.  For each source, we determine a maximum level, i.e. the level of the most deeply nested structure. If the maximum level of a region is relatively high, the region exhibits a more complex hierarchical structure. 

\begin{figure}[htb!] 
\plotone{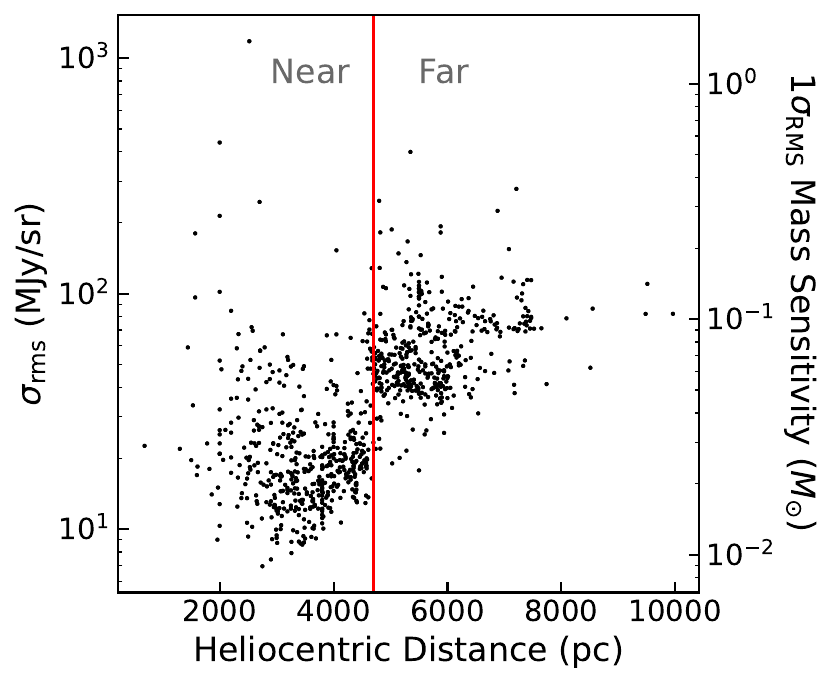}
\caption{Scatter plot showing the difference in the individual RMS values ($\sigma_{\text{rms}}$) and the corresponding 1$\sigma_{\text{rms}}$ point source mass sensitivity for near and far sample targets within the sub-sample of 900 ALMAGAL targets with corresponding clump property measurements. The  $\sigma_{\text{rms}}$ values correspond to the standard deviation of the residual images after masking out regions of bright emission in the original (non-smoothed) intensity images \citep[see][for more details]{Sanchez-Monge_2025}.  The red, vertical line indicates $d = 4.7$ kpc, the heliocentric distance dividing the near and far sample targets.} 
\label{fig:rms_dist}

\end{figure}

Using the distribution of the maximum dendrogram level for targets across the sample, we can determine appropriate thresholds for classification. As seen in Figure \ref{fig:max_lvl_dists}, the distribution for the maximum level for each clump is continuous, although distinct in shape and extent for the near and far samples. This makes sense, even though the original, native resolution ALMAGAL images have a largely uniform flux density sensitivity in Jy/beam, the distribution of beam-independent RMS values ($\sigma_{\text{rms}}$) in units of MJy/sr varies between the near and far sample, as shown in Figure \ref{fig:rms_dist}. This is directly related to the difference in mass sensitivity between the two samples, which ultimately affects the number of levels in the output dendrograms, as we are more sensitive to lower mass structures in the near sample than in the far sample. 

Since the dendrogram parameters $f_{\text{min}}$ and $\delta$ are scaled by the individual $\sigma_{\text{rms}}$ of each target, setting the same maximum level threshold will result in a distance bias where sources in the near sample are disproportionately flagged as `Rich', and far sample sources are more likely to be flagged as `Isolated'. To avoid this distance bias, we impose different maximum level thresholds for the near and far sample based on their individual distributions.

With this method, we are able to define four populations of clumps in our sample: Empty, Isolated, Simple, and Rich. The Empty sub-sample is composed of clumps without dendrogram detections in the hierarchical structure catalog. Isolated targets are minimally fragmented, Simple targets have an intermediate level of sub-structure, and Rich targets have the highest level of morphological hierarchy. We provide a few representative examples from our sample for each population in Figure \ref{fig:clump_class}. 

We note that there are 48 Empty targets that have core detections from the ALMAGAL core catalog \citep{Coletta_2025} but no dendrogram detections. Additionally, there are 10 targets that have dendrogram detections, but no core detections, all of which are part of the Isolated sub-sample. We attribute this discrepancy to the difference in the data products and the structure identification algorithm used for this catalog and the ALMAGAL core catalog. For this catalog we use a spatially smoothed, common linear resolution version of the ALMAGAL data and a dendrogram-based extraction procedure, whereas the ALMAGAL core catalog is generated using the native resolution data products with the \textit{CuTEx} algorithm. Additionally, for the hierarchical structure catalog, a structure is required to have a peak intensity of $3\sigma_{\text{rms}}$ and a minimum intensity threshold of $2\sigma_{\text{rms}}$ at the structure boundary. For the ALMAGAL core catalog, core detections must have a peak intensity of at least $5\sigma_{\text{rms}}$ and there is no minimum intensity threshold imposed.

In Figure \ref{fig:max_lvl_dists}, we show the exponential distribution fits for the maximum dendrogram levels found in the near and far samples, and indicate the $1\sigma_{\text{level}}$ and $3\sigma_{\text{level}}$ for each distribution. Since the maximum dendrogram level is always an integer value, we round these values to the nearest integer. For their respective near and far distributions, all Isolated targets have a maximum level $<1\sigma_{\text{level}}$, Simple targets have a maximum level $\geq1\sigma_{\text{level}}$ and $<3\sigma_{\text{level}}$, and Rich targets have a maximum level $>3\sigma_{\text{level}}$. The corresponding integer ranges are shown in the panels of Figure \ref{fig:max_lvl_dists}. We summarize the categorical cross-sections of our ALMAGAL sample in Table \ref{tab:sample}. 

\subsection{Calculation of Continuum Structure Properties}
\label{subsec:struc_prop}

For each continuum structure in the catalog, we report its position, effective radius $R_{\text{eff}}$, integrated flux $F_{\nu}$, and peak intensity $I_{\text{peak}}$, as shown in Table \ref{tab:catalog}.  The coordinates of the structure correspond to the average position of the continuum structure, as measured within the dendrogram contour. We calculate an effective projected radius from the total structure area: $R_{\text{eff}} \equiv (N_{\text{pix}} A_{\text{pix}} / \pi)^{1/2}$, where $N_{\text{pix}}$ is the total number of pixels in the structure, and $A_{\text{pix}}$ is the physical area of each pixel in au$^{2}$. The integrated flux of the structure $F_{\nu}$ is obtained by summing the pixel values in the primary beam corrected image within the dendrogram structure and converting to Jy. For non-trunk structures we use the ``clipping paradigm" (see \citealt{Rosolowsky_2008}) to calculate the background subtracted integrated flux $F_{\nu \text{,bgs}}$ by subtracting the minimum intensity of the structure from each pixel and then summing over the entire structure. 

Assuming optically thin dust continuum emission, we derive an isothermal mass estimate for each structure using 
\begin{equation}
    M_{\text{struc}} = \frac{d^2 F_{\nu} R_{\text{gd}}}{\kappa_{\nu} B_{\nu}(T_{d})}
\label{mass_eq}
\end{equation}
where $d$ is the distance to the ALMAGAL source from \cite{Molinari_2025}, $F_{\nu}$ is the integrated flux of the structure, $R_{gd}$ is the gas-to-dust ratio,  $\kappa_{\nu}$ is the dust opacity per unit mass, and $B_{\nu}(T_{d})$ is the Planck function at the local dust temperature $T_d$. We take the gas-to-dust ratio to be 100 and we use a dust opacity per unit mass of $\kappa_{\text{1.3mm}} = 0.9~$cm$^{2}$g$^{-1}$ based on the \cite{Ossenkopf_1994} model for moderately coagulated dust grains with thin ice mantles at gas densities $\geq 10^6$cm$^{-3}$ and a coagulation timescale of 10$^5$ yr. We also calculate a background subtracted mass using the background subtracted integrated flux of the structure, $F_{\nu,bgs}$. 

Ideally, the gas temperature of each structure would be measured using spectral line emission from high density molecular tracers, and assuming the gas and dust are coupled, we could use that for the dust temperature. However, we consider an extensive analysis of the temperature of the full hierarchy of continuum structures for the entire ALMAGAL sample to be beyond the scope of this work. For the purposes of this paper, we adopt the same temperature assumptions used for the ALMAGAL core catalog \citep{Coletta_2025} and for cores extracted from the 7M+TM2 continuum data in \cite{Wells_2024}. This model assigns core temperatures according to its evolutionary stage based on the calibration of the clump $L/M$ ($L_{\odot}/M_{\odot}$) with the rotational temperature CH$_3$CCH performed in \cite{Molinari_2016a}:

 \[ T = \begin{cases} \mbox{20 K} & \mbox{if } L/M<1 , \\ 
 \mbox{35 K} & \mbox{if }  1  \leq \textit{L/M} < 10 ,\\
  21.1(L/M)^{0.22} & \mbox{if }  \textit{L/M} \geq 10 .
 \end{cases} \]

Additionally, we calculate the mass surface density by dividing the mass by the area of the structure:

\begin{equation}
    \Sigma_{\text{struc}} = \frac{M_{\text{struc}}}{N_{\text{pix}}A_{\text{pix}}}.
\end{equation}

For trunk and branch structures, we include the mass of the \textit{entire} structure, including all sub-scale branch and leaf sub-structures. 

In the analysis discussed in Section 5.2, we calculate a local surface density ($\Sigma_{\text{surr}}$) for each core identified in the ALMAGAL compact core catalog \citep{Coletta_2025}. This “surrounding density” is computed in Equation 3 where $M_{\text{struc}}$ is the mass enclosed within the smallest dendrogram structure that contains the core and also has an effective radius greater than 3 times the core radius. This restriction ensures that this surface density describes the environment immediately surrounding the core, and not the core itself.

The surrounding surface density $\Sigma_{\text{surr}}$ is calculated by subtracting the total non-background subtracted mass of cores, $M_{\text{cores}}$, contained within a dendrogram structure from the mass of the dendrogram structure, $M_{\text{struc}}$ 

\begin{equation}
    \Sigma_{\text{surr}} = \frac{(M_{\text{struc}}-\sum_{i=0}^{k}M_{\text{k,core}})}{\pi R_{\text{eff}}^2}.
\end{equation}

\begin{figure}[htb!]
\epsscale{1.0}

\plotone{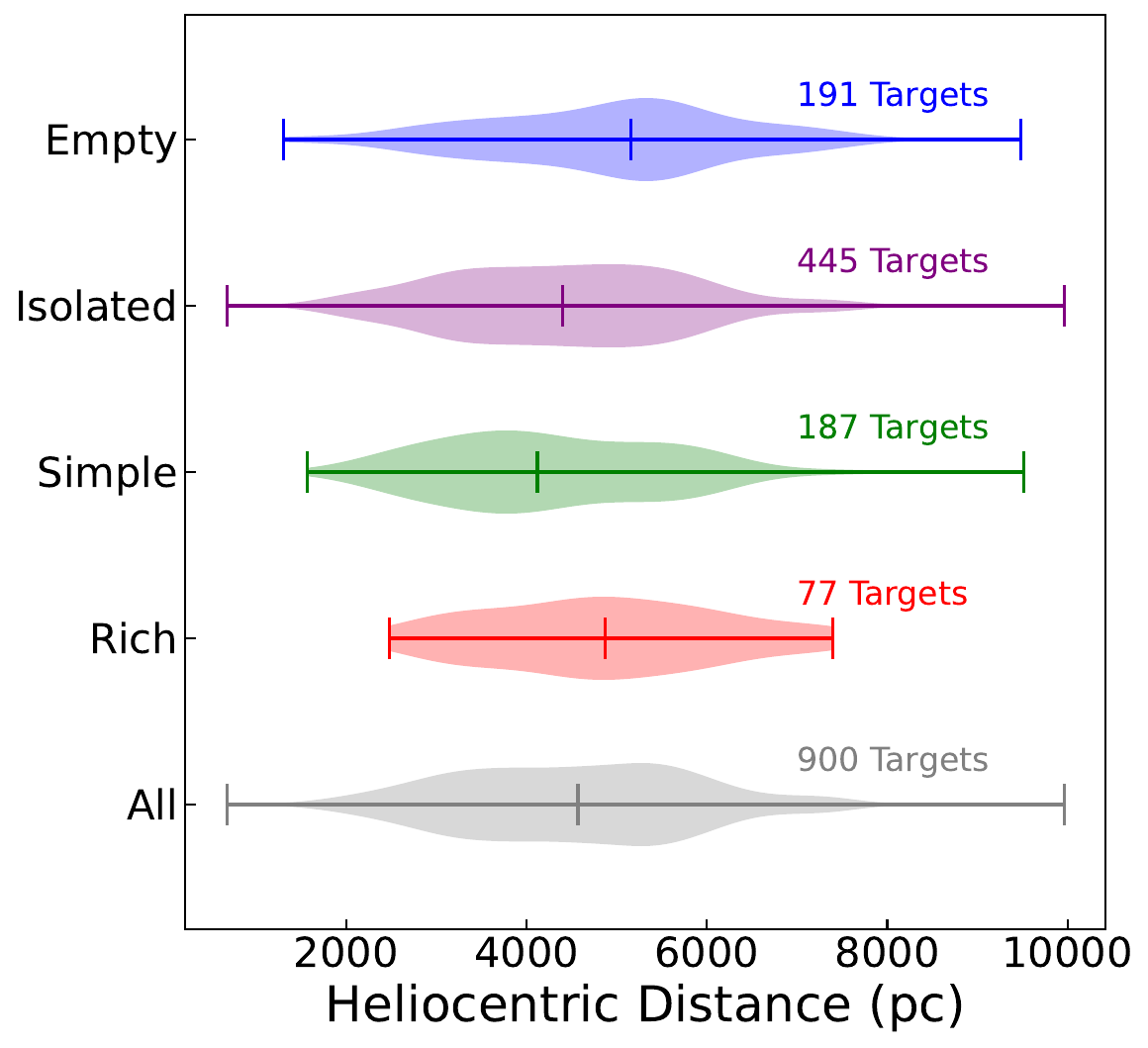}
\plotone{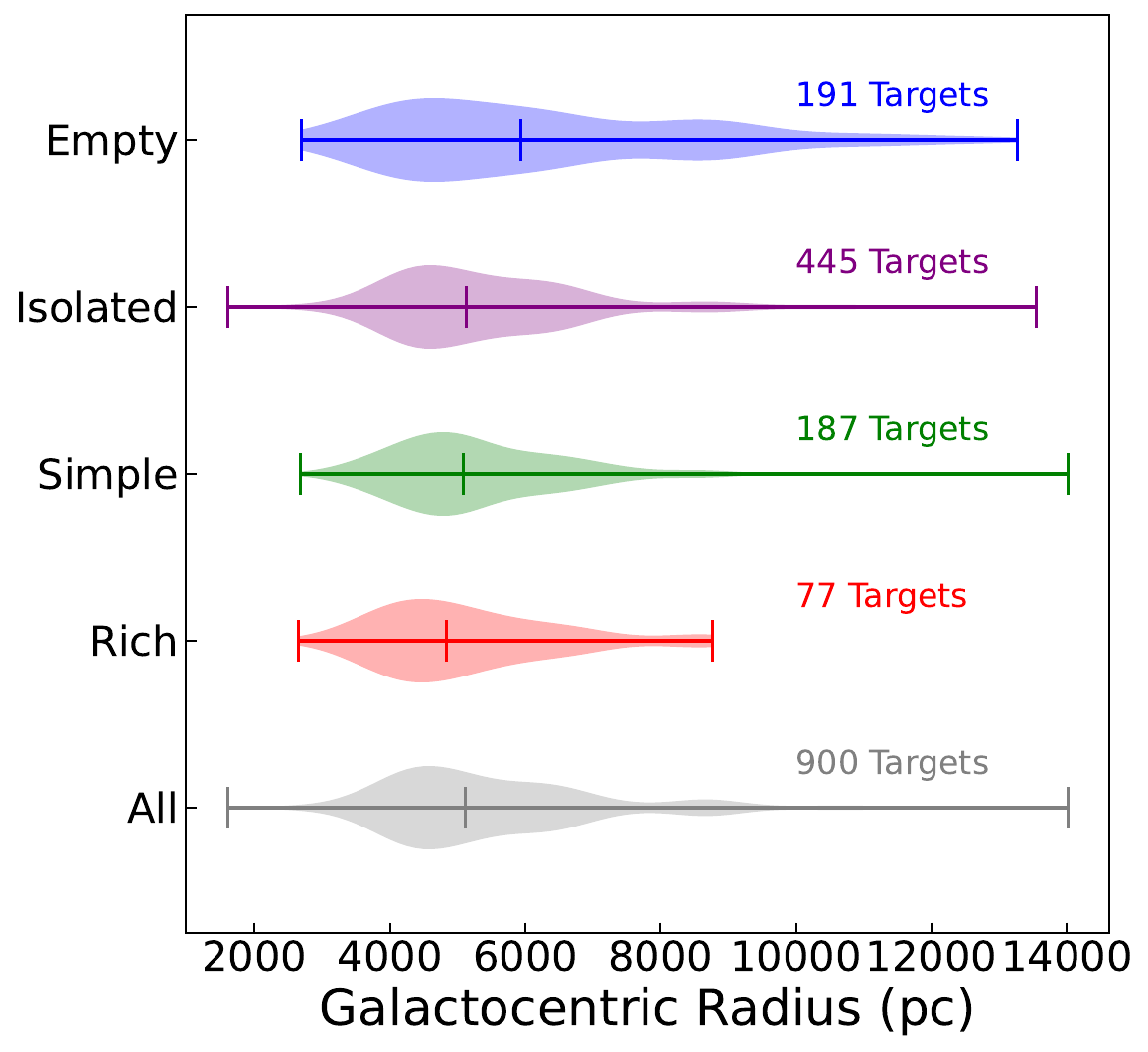}
\caption{Violin plots showing the distribution of heliocentric distances (Top) and Galactocentric radii (Bottom) for the different morphological categories of clumps in the ALMAGAL sample (Empty, Isolated, Simple, Rich and All). The center vertical line indicates the median for each distribution.} 
\label{fig:dist_hist}

\end{figure}

\section{Results}
\label{sec:results}

We present the distributions of various clump properties as they relate to the hierarchical morphology observed in our ALMA data. 
We look at 4 subcategories in our sample covering different levels of fragmentation: Empty, Isolated, Simple, and Rich. The Isolated, Simple, and Rich targets are defined based on the method outlined in Section \ref{subsec:clump_class}.

We note that 1007 of the 1013 ALMAGAL clumps have measured physical properties (Elia et al. 2025, submitted). For our sub-sample, 900 of the 904 total targets have measured physical properties.

\subsection{Evaluating Distance Bias}
\label{subsec:dist_bias}

In Figure \ref{fig:dist_hist}, we present the heliocentric distance and Galactocentric radius distributions for each morphological sub-sample.
We see that these distance distributions are similar for the Isolated, Simple, and Rich samples -- however the Empty regions have a bias towards larger distances. This makes sense since the Isolated, Simple, and Rich samples are defined using distribution dependent thresholds on their maximum dendrogram level. However, the Empty targets include only non-detections, meaning that this category includes sources for which there was no significant emission above a certain noise-based threshold. Since the far sample ($>4.7$ kpc) targets in general exhibit higher $\sigma_{\text{rms}}$ values (see Figure \ref{fig:rms_dist}), we expect the Empty sample distribution to be overrepresented at larger distances. We note that the Rich targets have a slight overabundance at greater heliocentric distances, but this is likely due to the larger spatial field-of-view present in targets located further away.

\subsection{Comparing Hierarchical Morphology and Clump Properties}
\label{subsec:clump_distributions}

In Figure 5 we present violin plots showing the distributions of clump physical properties for the full sample as well as for the different morphological sub-samples we define in Section \ref{subsec:clump_class}. We find that targets with more complex hierarchical structure tend to have distributions containing higher surface densities, warmer dust temperatures, and larger \textit{L/M} ratios at the clump-scale. The Rich sample has a median mass surface density of 1.28 g cm$^{-2}$, the highest out of the different morphological sub-categories. The median surface density continues to decrease as we move from the Rich to the Empty sub-samples, where Empty targets have a median surface density of 0.35 g cm$^{-2}$.

The Rich sub-sample contains the clumps with the highest dust temperatures (up to 39.5 K), with a median $T_{\text{dust}}$ $\sim24$~K. This is a $\geq10$ K offset from the median $T_{\text{dust}}$ of the Empty and Isolated sub-samples, with values of 12.2$-$13.2 K, respectively. 

\begin{figure*}[htb!] 
\epsscale{0.5}
\begin{centering}
\plotone{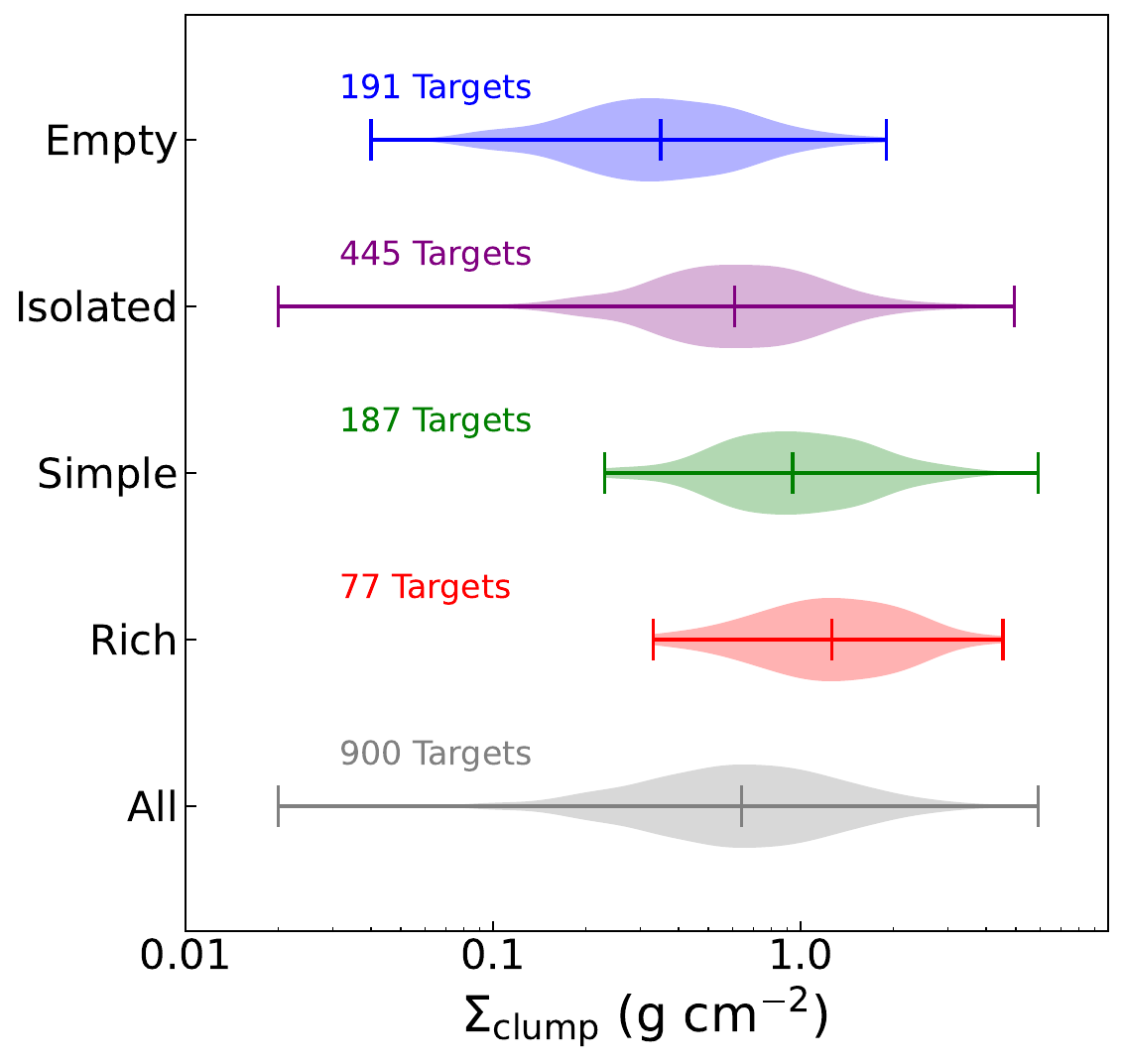}
\plotone{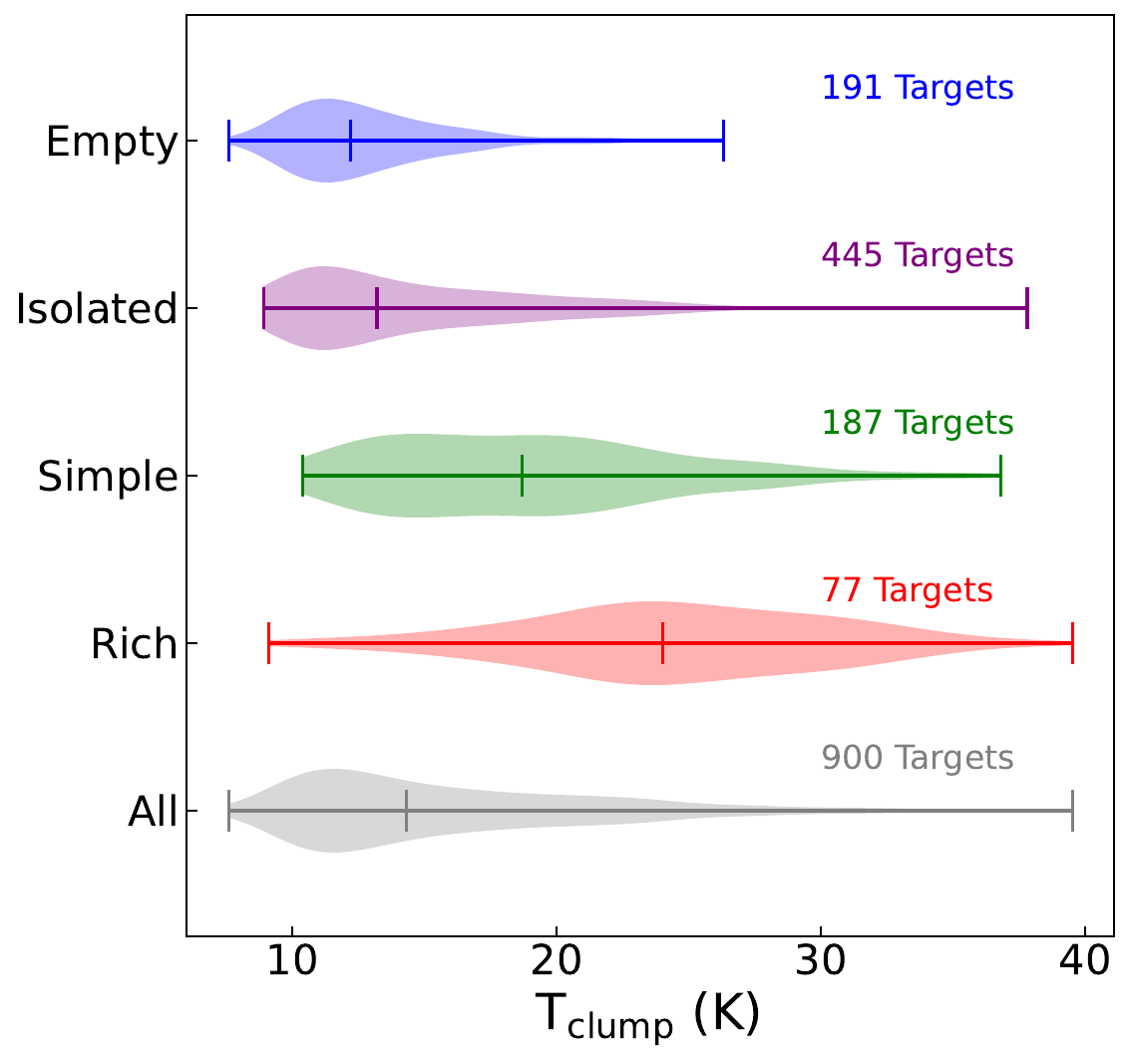}
\plotone{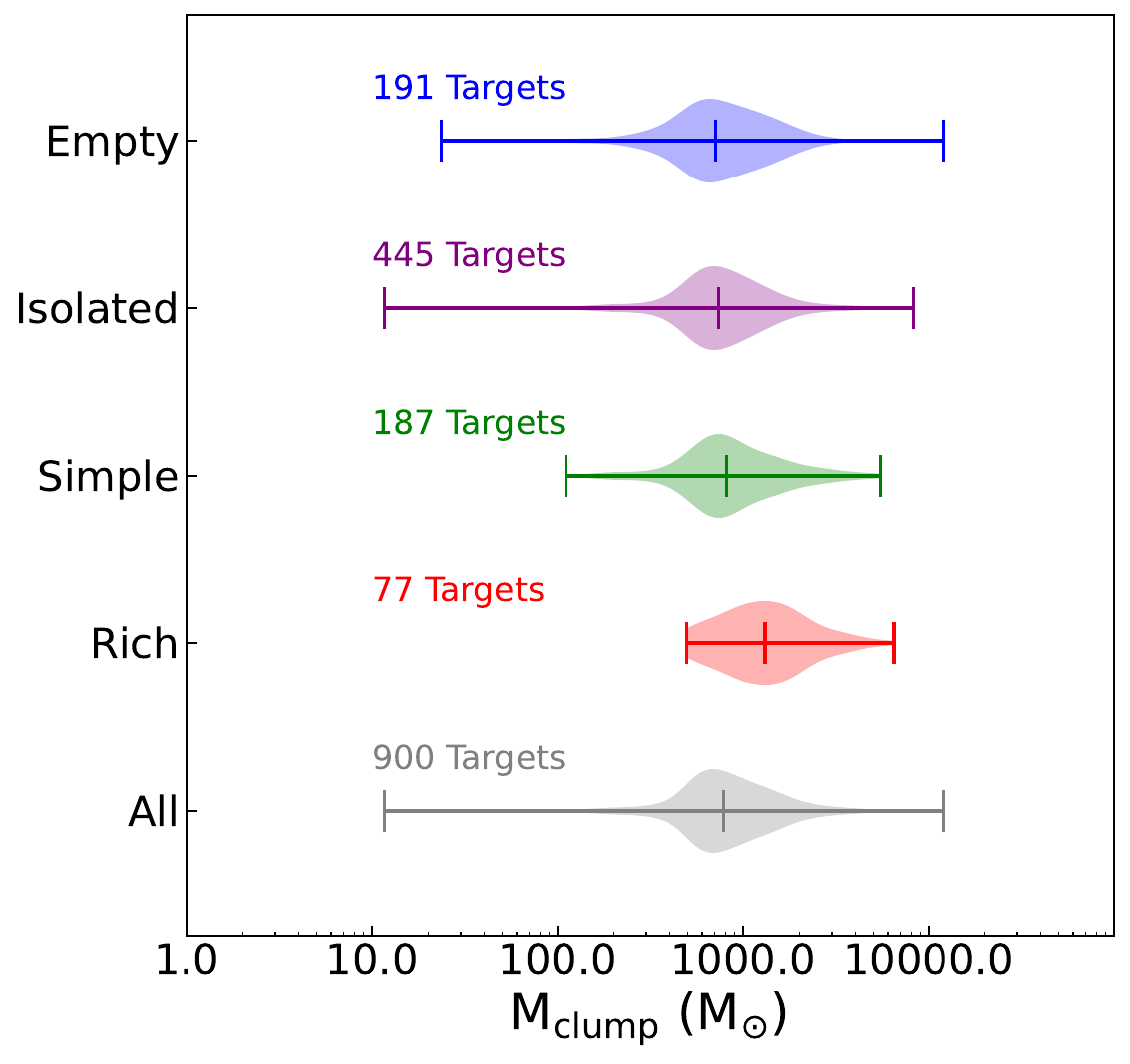}
\plotone{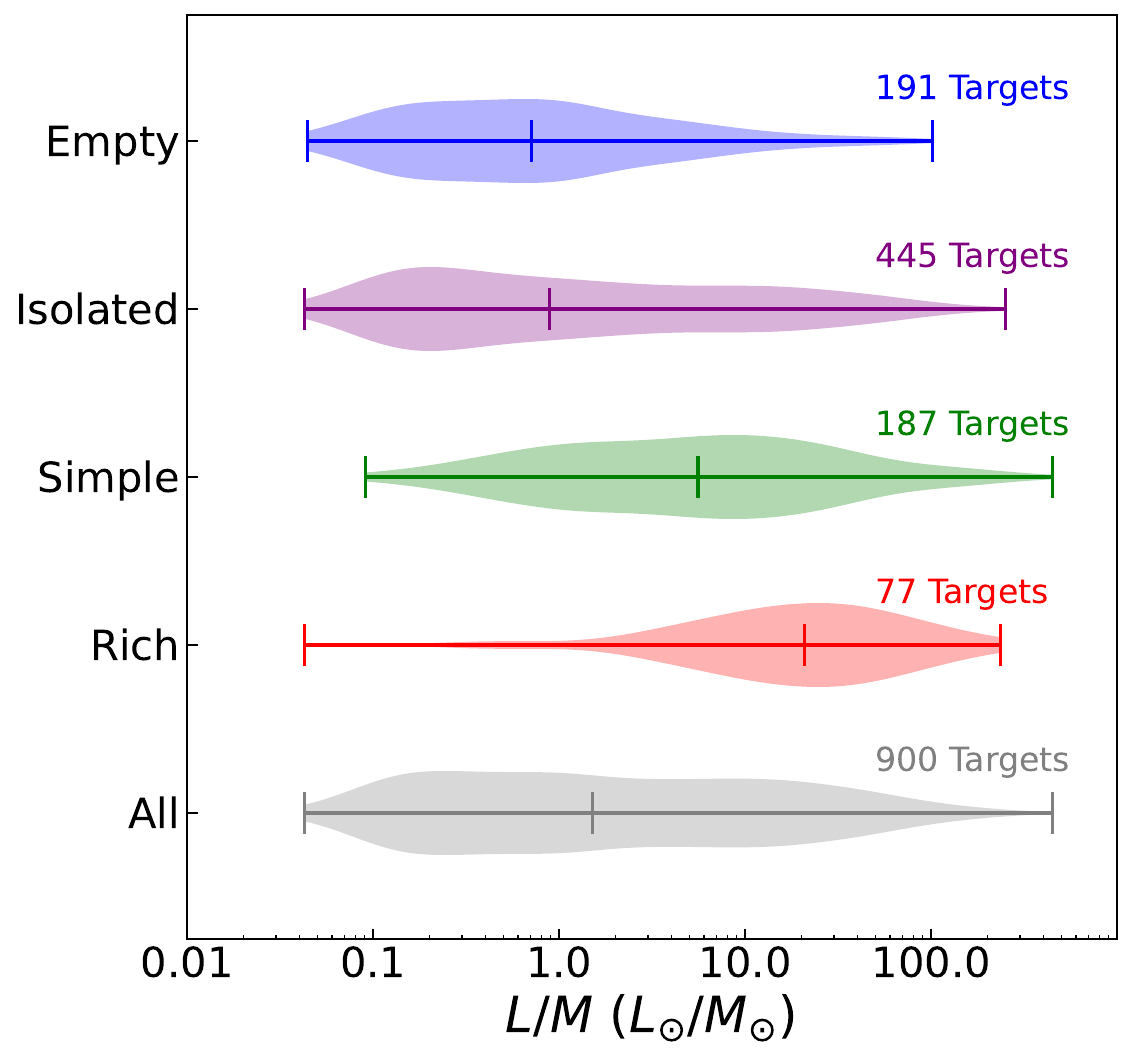}
\plotone{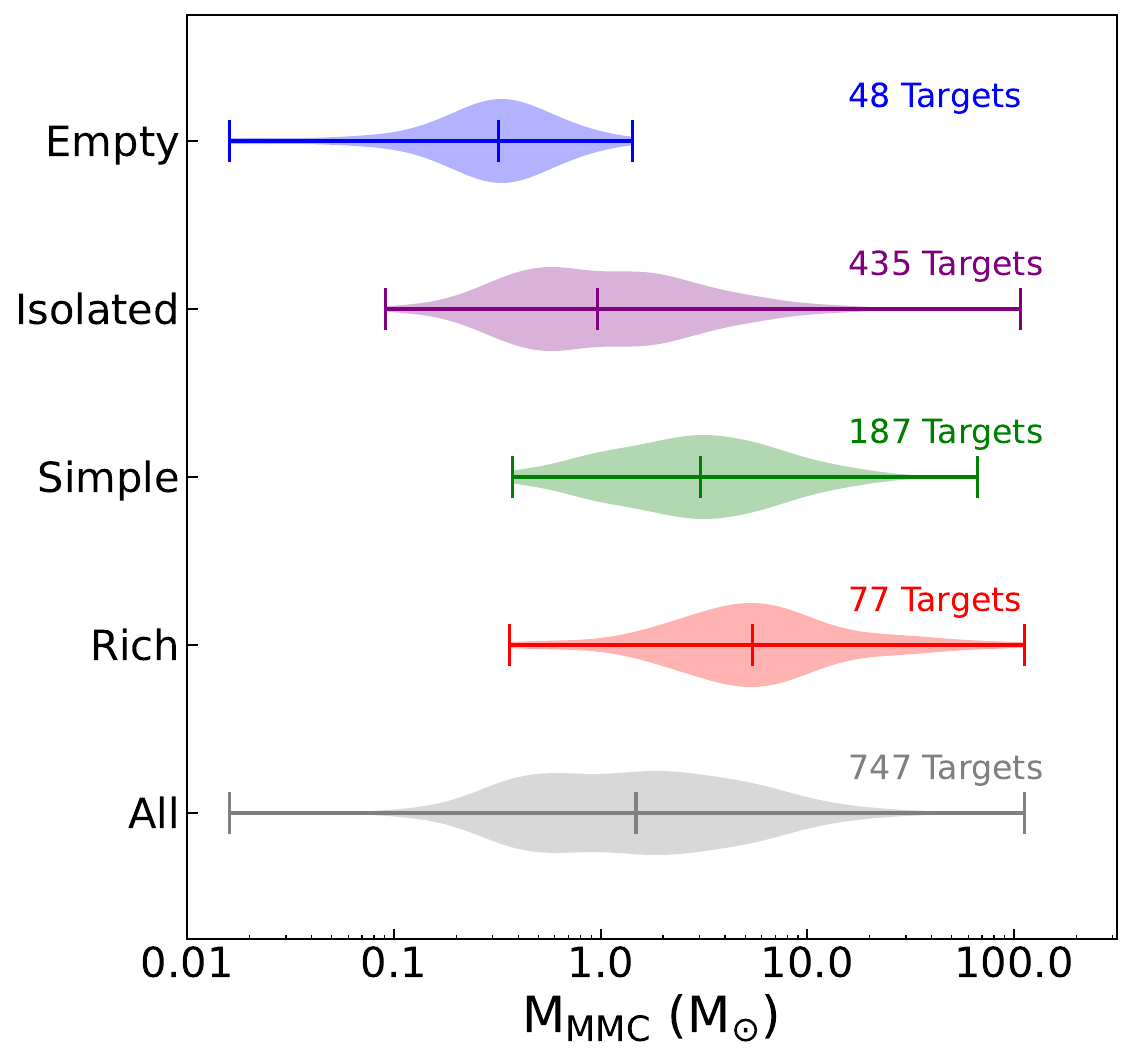}
\plotone{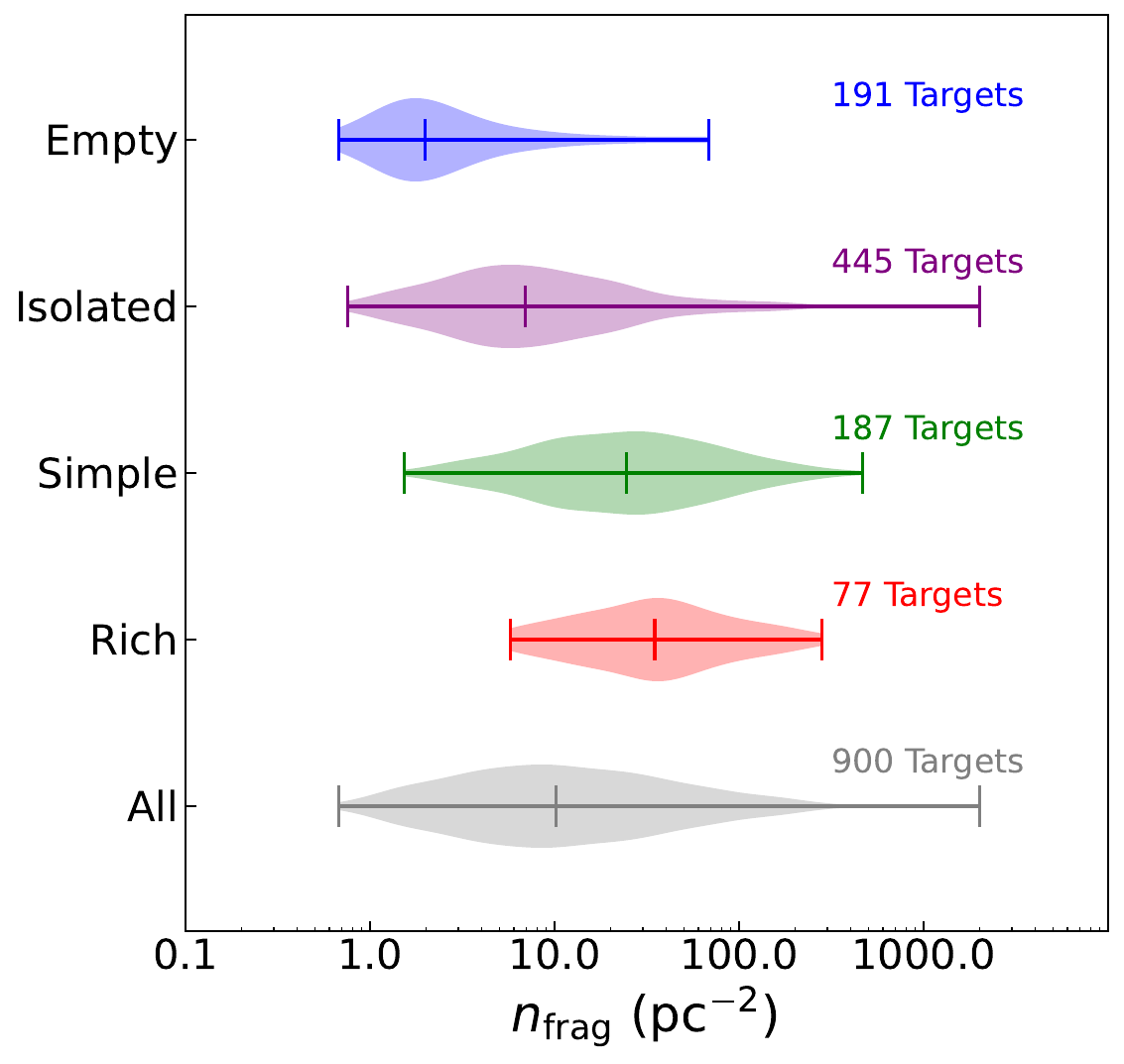}
\caption{Violin plots highlighting the population differences in clump surface density (Top Left), dust temperature (Top Right), mass ($M_{\text{clump}}$) (Middle Left) \textit{L/M} ratio (Middle Right), MMC mass (Bottom Left), and core number density ($n_{\text{frag}}$) across the Empty, Isolated, Simple, and Rich clump sub-samples as well as the full sample. The center vertical line indicates the median for each distribution. In general, the morphologically rich targets have a higher median surface density, dust temperature, \textit{L/M}, MMC mass, and core number density, all of which are properties that may be related to the evolutionary stage of the clump. The mass is the only clump property that does not vary significantly between the different mophological subsamples.}
\label{fig:clump_prop}
\end{centering}
\end{figure*}

The vast majority (93.3\%) of Rich targets have an \textit{L/M} ratio greater than $1~L_{\odot}/M_{\odot}$. On the other hand, 6.28\% of the Empty sample does not exceed an \textit{L/M} ratio of $10~L_{\odot}/M_{\odot}$. Considering that the clump mass ($M_{\text{clump}}$) distribution  does not significantly change between sub-samples (Figures \ref{fig:dist_hist} and \ref{fig:ks_tests}), the increasing $L/M$ with hierarchical complexity can be largely attributed to an increase in $L_{\text{clump}}$. 

In general, the Empty and Isolated samples have lower MMC masses than the Simple and Rich targets, as seen in Figure \ref{fig:clump_prop}. For clumps with detections in the ALMAGAL core catalog, the median MMC mass is $\sim 0.3 ~M_{\odot}$ for Empty regions, $\sim 1~M_{\odot}$ for Isolated regions, $\sim3~M_{\odot}$ for Simple regions, and $\sim 5~M_{\odot}$ for Rich regions. There are no MMCs with masses below $0.3 ~M_{\odot}$ for Simple or Rich targets. 

We provide a more detailed summary of the statistical quantities for these distributions in Table \ref{clump_stat_tab} (located in Appendix \ref{clump_stat}.)

\subsection{K-S Tests}
\label{subsec:ks_tests}

\begin{figure}[htb!] 
\begin{centering}
\epsscale{1.0}
\plotone{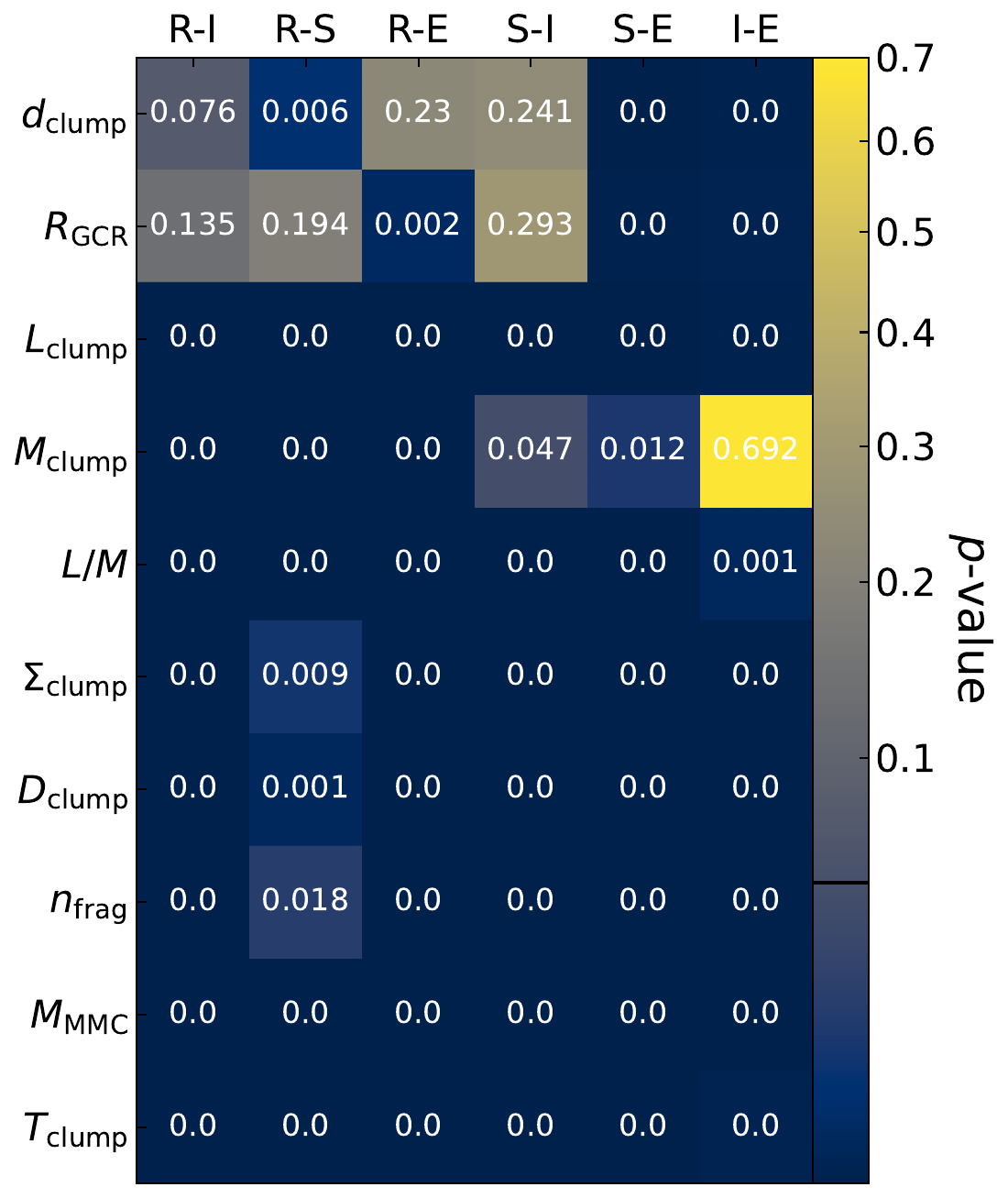}
\caption{Heatmap matrix for Two Sample K-S Test \textit{p}-values for each combination of the Rich (R), Simple (S), Isolated (I), and Empty (E) source populations. All \textit{p}-values are shown in white text. A \textit{p}-value $= 0.05$ is shown with a solid black line on the color bar. The \textit{p}-values $< 0.05$ indicate combinations where we reject the null hypothesis that both samples are pulled from an identical distribution at a confidence level of 95\%. The clump properties are listed on the $y$-axis, where $d_{\text{clump}}$ is the heliocentric distance, $R_{\text{GCR}}$ is the Galactocentric radius, $L_{\text{clump}}$ is the clump luminosity, $M_{\text{clump}}$ is the clump mass, $L/M$ is the clump luminosity-to-mass ratio, $\Sigma_{\text{clump}}$ is the clump surface density, $D_{\text{clump}}$ is the clump diameter, $n_{\text{frag}}$ is the fragment density of cores from the ALMAGAL core catalog \citep{Coletta_2025} in the clump (the number of cores divided by the area of the observed field of view), and $T_{\text{clump}}$ is the clump dust temperature.} 
\label{fig:ks_tests}
\end{centering}
\end{figure}

We quantitatively determine if the clump property distributions for the different sub-categories (Rich, Simple, Isolated, Empty) are statistically distinct using a two sample Kolmogorov-Smirnov (K-S) test. The null hypothesis for this test is that the samples from the two data sets come from the same underlying distribution. The K-S test produces a probability values (\textit{p}-value) that is compared to a chosen significance level (most commonly 0.05). We use the \verb|kstest| function from the Python package \verb|scipy|. A summary of the K-S test of each clump property for each sub-population category is given in Figure \ref{fig:ks_tests}. 

Most sub-sample combinations have statistically similar distributions for heliocentric distance and Galactocentric radius. The exception to this is the Empty clump sub-sample, which has distance and Galactocentric radius distributions that are distinct from the other sub-samples, likely because of the lower dendrogram detection rate in `Far' sample sources as discussed in Section \ref{subsec:dist_bias}. We also note that the Rich and Simple distributions have statistically dissimilar distributions for heliocentric distance at the 95\% confidence level. This is likely due to rounding our maximum dendrogram level thresholds to the nearest integer for our morphological classifications, which causes a lack of precision in ensuring a perfectly even sampling of morphological groups between the near and far sample. 

All other morphological sub-categories have clump property distributions that are statistically distinct according to our K-S test calculations, aside from the clump masses, which appear to be drawn from the same distribution for the Empty and Isolated sub-samples only.

\subsection{Structure Properties}
\label{subsec:res_struc_props}

\begin{figure*}[htb!] 
\begin{centering}
\epsscale{1.1}
\plottwo{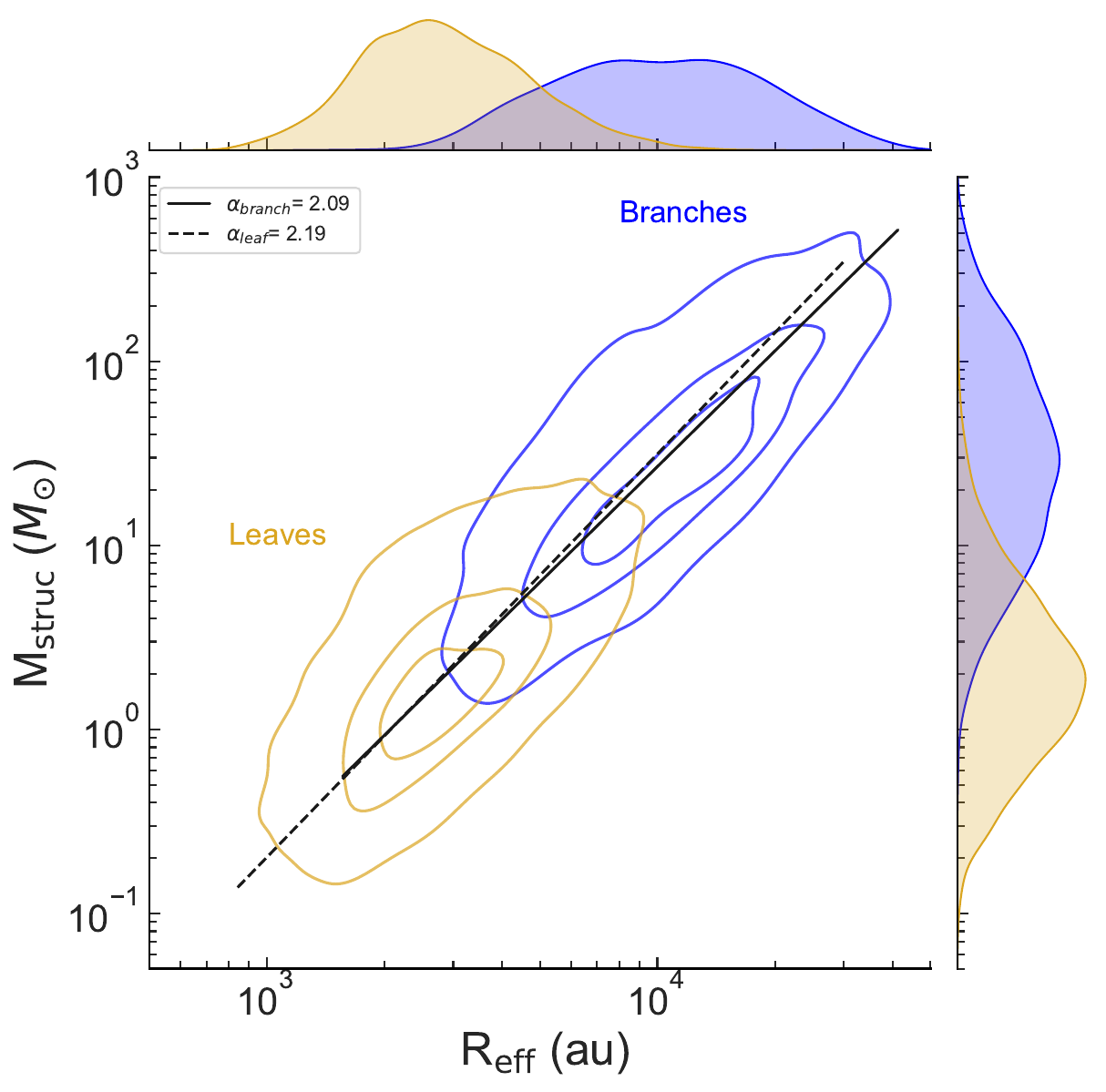}{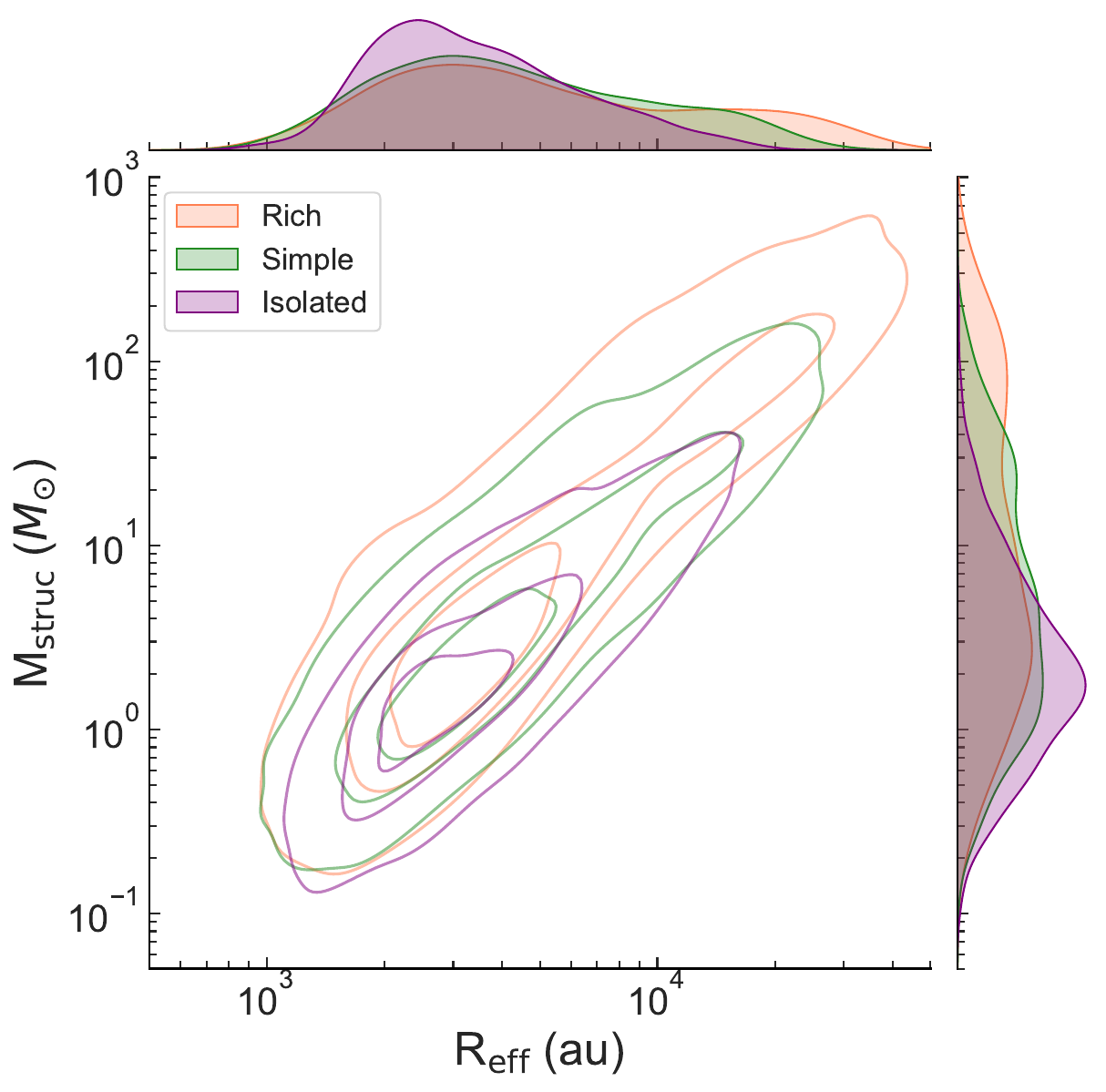}
\caption{(Left) A 2D Kernel density estimate plot of the mass-radius relation for leaf (green) and branch structures (red) in the ALMAGAL hierarchical structure catalog. Best fits corresponding to the power law relation $M \propto R^{\alpha}$ for each population are indicated with the solid and dashed black lines. (Right) A 2D kernel density estimate plot for the mass-radius relation for all dendrogram structures in the Rich (orange), Simple (green), and Isolated (purple) sub-samples. For both panels, the contours indicate the shape of the distributions for the inner [90, 50, 20]\% of the data} 
\label{fig:mvr_full}
\end{centering}
\end{figure*}

We present the relationship between the mass and radii for structures in our catalog in Figure \ref{fig:mvr_full}. In the left panel, we examine the branch and leaf populations separately, and in the right panel we compare the Rich, Simple and Isolated populations. As seen in both panels of Figure \ref{fig:mvr_full}, the structures in our hierarchical catalog have effective radii ($R_{\text{eff}}$) spanning $\sim 840-42000$ au. Leaf structures span the range of $840-30100$ au, with a median value of $\sim 2700$ au. It should be emphasized that leaf structures are not analogous to the cores identified in the ALMAGAL compact core catalog, as this range in radii exceeds the range in compact core radii, with R$_{\text{core}}$ between $\sim 400-1500$ au \citep{Coletta_2025}. We note that it is possible to have leaf structures with R$_{\text{eff}}$ somewhat smaller than 1000 au, which is 1/2 the linear beam size of 2000 au. This is due to our choice in $n_{\text{pix}} = 0.5\times A_{\text{beam,pix}}$ when generating the dendrogram, resulting in some structures that are slightly smaller than the beam. Branch structures have $R_{\text{eff}}$ covering $~1500-42000$ au, with a median value of $\sim 10000$ au. 

The structures in our hierarchical catalog have masses ranging from $0.05$ to $670 ~M_{\odot}$, with a median mass of 3.6 $M_{\odot}$. Leaf structures have masses ranging from $0.05-310 ~M_{\odot}$ \footnote{Excluding a single outlier leaf structure with $M = 670.3 ~M_{\odot}$.}, with a median of 1.9 $M_{\odot}$. The larger branch structures have a mass range of $\sim 0.6-660 ~M_{\odot}$, with a median value of 27.4 $M_{\odot}$. We note that the $M \propto R^{\alpha}$ relation for leaf and branch structures is nearly the same, with the estimates for $\alpha$ being 2.21 and 2.08, respectively (Figure \ref{fig:mvr_full}).   

In the right panel of Figure \ref{fig:mvr_full}, we show how the mass and radii of our structures vary as a function of their morphological sub-category. All three groups of targets follow the same trend, with regions exhibiting more hierarchical structure containing the largest and most massive structures. For the Rich, Simple, and Isolated sub-categories, the median masses of structures are $\sim$[6.8, 4.1, 2.0] $M_{\odot}$, and the median effective radii are $\sim$[4290, 3940, 3150] au, respectively. 

In Appendix \ref{sec:near_far_mvr_diff}, we evaluate structure property differences between the near and far sample, and find that both samples extract structures of similar size and mass. We note that the near sample observations have more low-mass structures, but this is to be expected, as the far sample has a poorer mass sensitivity.

\begin{figure*}[htb!] 
\begin{centering}
\epsscale{1.1}
\plottwo{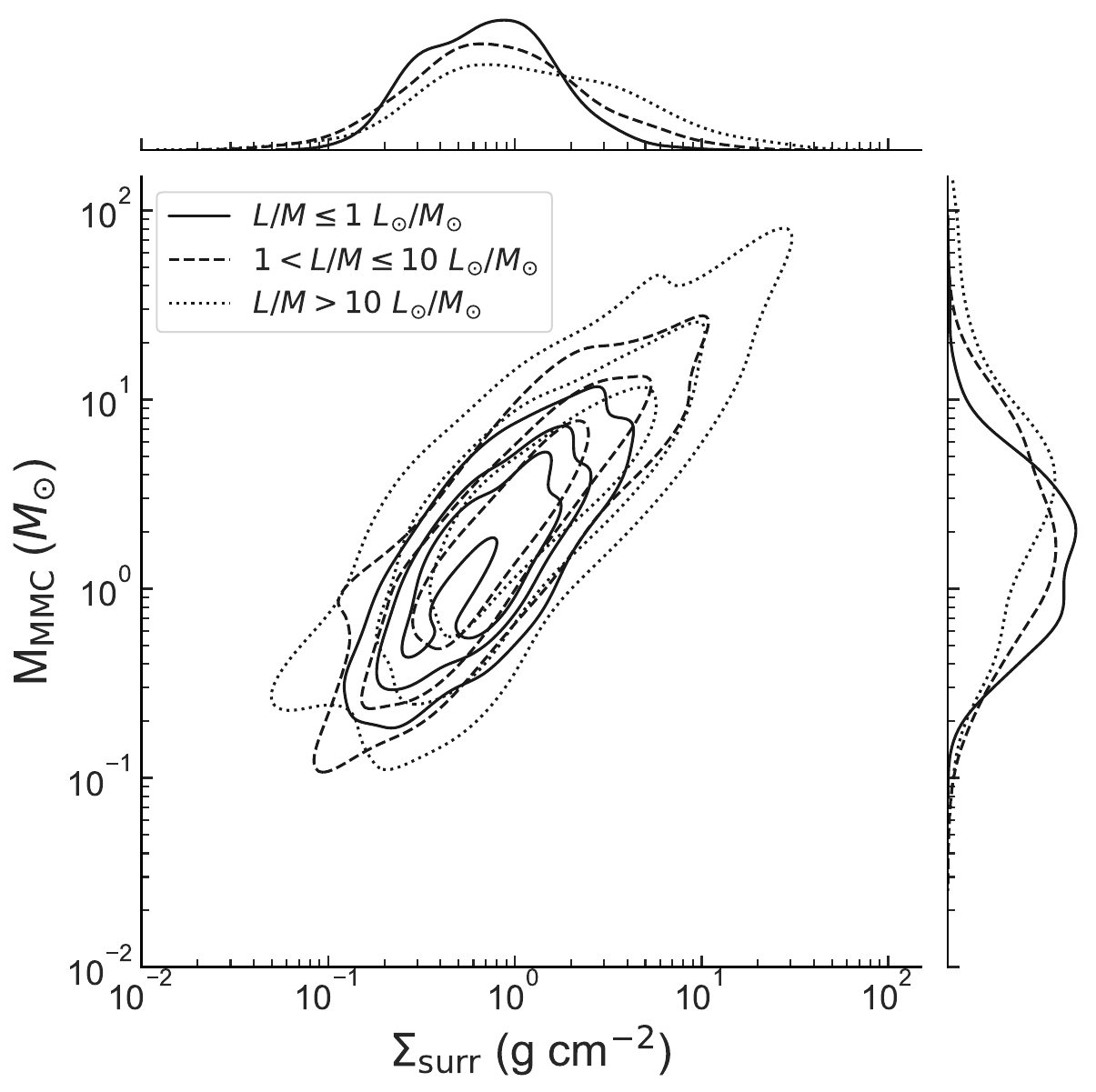}{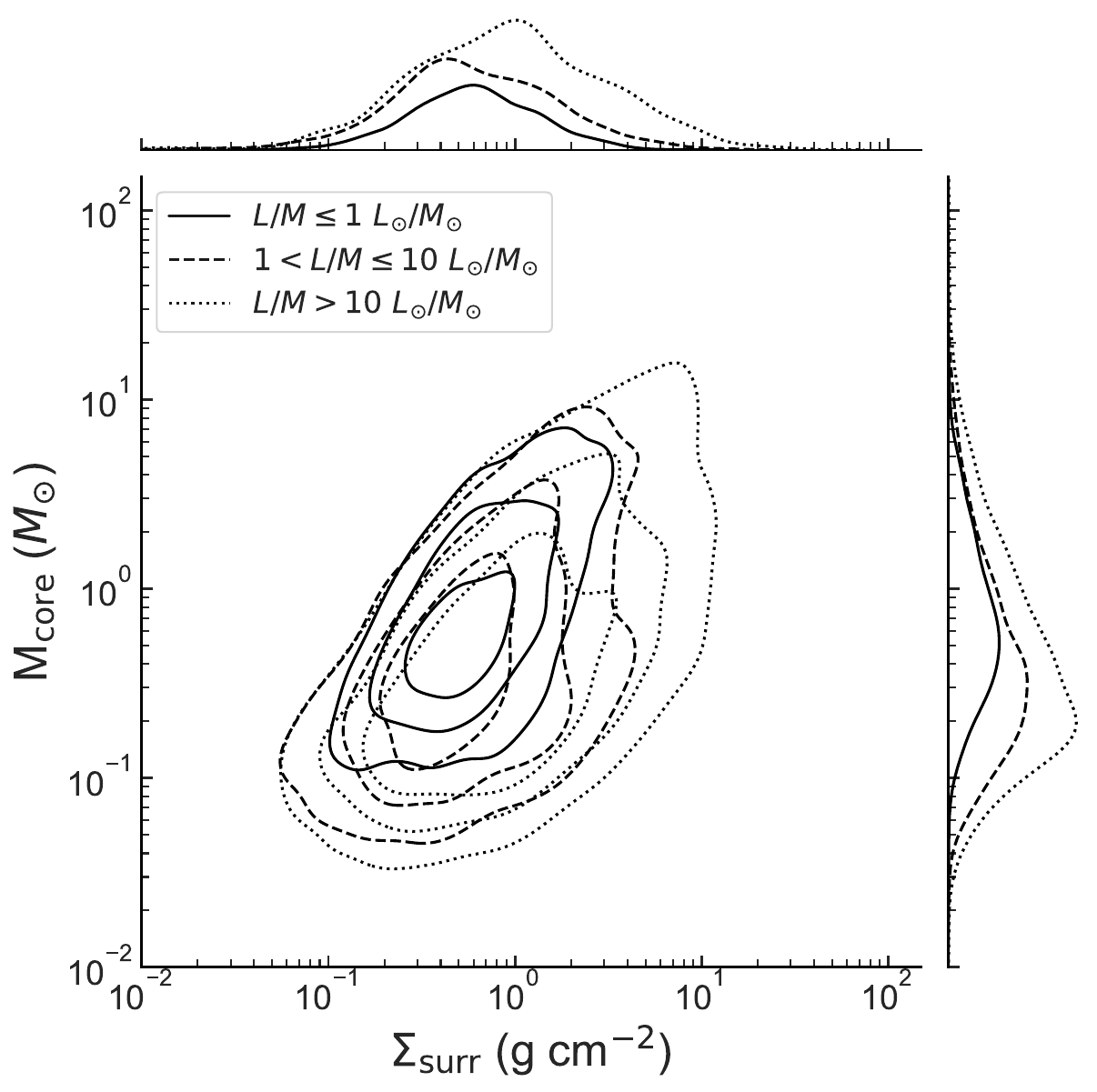}
\caption{(Left panel) A 2D kernel density estimate plot of the mass of the MMC as a function of the surface density of its most immediate underlying dendrogram structure that has an effective radius at least 3 times the radius of the core ($\Sigma_{\text{surr}}$). (Right panel) The same figure, but we include all cores from the ALMAGAL core catalog. ALMAGAL targets are divided into three evolutionary categories based on their \textit{L/M} ratio.  Contours indicate the shape of the distribution for the inner [90, 70, 40]\% of the data.  Marginal plots indicate the 1D kernel density estimates of the dendrogram structure surface density and core mass on the \textit{x}- and \textit{y}-axes, respectively. }
\label{fig:surfd_mcore}
\end{centering}
\end{figure*}

\section{Discussion}
\label{sec:discussion}

\subsection{Hierarchical Structure and Clump Evolution}
\label{subsec:morph_evol}

In Section \ref{subsec:clump_class}, we examined various clump evolutionary indicators, such as dust temperature, surface density, and MMC mass, and compared their distributions for our different morphological sub-samples (Empty, Isolated, Simple, and Rich). In general, we find that clumps exhibiting an increased level of hierarchical complexity have distributions shifted towards higher dust temperatures, surface densities and MMC masses. The KS-tests performed in Section \ref{subsec:ks_tests} confirm that our morphological sub-samples have distinct distributions in most of the properties surveyed, with significant ($p<0.05$) evidence they do not draw from the same underlying distribution. ALMAGAL targets from the Rich sub-sample show the most structural hierarchy and have an \textit{L/M} ratio distribution that heavily skews towards higher L/M, with a median value of 20.7 L$_{\odot}$/M$_{\odot}$. We consider clumps with \textit{L/M} $> 10$ L$_{\odot}$/M$_{\odot}$ to be at a later stage of evolution, undergoing active star formation with the possible presence of H{\sc ii} regions \citep{2025_Cesaroni}.

Additionally, as seen in Figure \ref{fig:clump_prop}, the median MMC masses, dust temperatures and deconvolved surface densities for each clump increases as regions exhibit more morphological complexity. This may be due to the increasing fragmentation of the clump as it evolves \citep{Coletta_2025}. Additionally, the amount of material that has been funneled down to the $\sim$1000s au scales that we can observe in ALMAGAL should be higher for more evolved regions as they have had more time to accrete material from the intra-clump gas reservoir, which may contribute to the increased richness of structure that we see towards those targets.

\subsection{Core Mass Accretion at Different Evolutionary Stages}
\label{subsec:core_mass_accretion}

We investigate the correlation between the MMC mass in a clump and its surrounding surface density, $\Sigma_{\text{surr}}$ (see Section \ref{subsec:struc_prop} for details) in the left panel of Figure \ref{fig:surfd_mcore}. We see that the MMC mass is directly related to the surrounding surface density in its environment, with MMCs existing only in proportionally dense environments. Different evolutionary stages follow a similar linear trend, with clumps at later evolutionary stages (higher $L/M$) exhibiting more massive MMCs in higher density environments. 

We extend this analysis of comparing the core mass with its surrounding surface density, $\Sigma_{\text{surr}}$, to all sources in the ALMAGAL core catalog for which there is a corresponding dendrogram structure, and present these results in the right panel of Figure \ref{fig:surfd_mcore}. When we include the entire core population, we find that the core mass has a looser correlation with its surrounding surface density than when we consider just the MMCs. For all three evolutionary populations shown in Figure \ref{fig:surfd_mcore}, the distribution of $M_{\text{core}}$ vs $\Sigma_{\text{surr}}$ covers a larger parameter space than the distribution of $M_{\text{MMC}}$ vs $\Sigma_{\text{surr}}$, with the inclusion of lower mass cores located in higher surrounding surface density environments. This spread in the distribution is exaggerated for more evolved regions, with the extent of the outermost contours in Figure \ref{fig:surfd_mcore} increasing by a factor of $\sim2$ for both $M_{\text{core}}$ and $\Sigma_{\text{surr}}$ between the $L/M\leq1~L_{\odot}/M_{\odot}$ and $L/M>10~L_{\odot}/M_{\odot}$ populations. 

One interpretation for this trend is that cores competitively accrete mass from an intra-clump reservoir of gas, with some cores accreting mass more efficiently than others depending on their location within the clump \citep{Bonnell_2001, Bonnell_Bate_2006}. In this competitive accretion scenario, cores that are formed later, or in a location outside of the central region of the clump will have a \textit{slower} mass accretion rate, even as more gas is funneled down from the larger scale reservoir. If we examine the right panel of Figure \ref{fig:surfd_mcore} in the context of core mass accretion throughout the evolution of a clump, we see that cores in clumps at early stages of their evolution have a tighter relationship between $M_{\text{core}}$ and $\Sigma_{\text{surr}}$. As this clump evolves,
certain cores will continue to accrete material and maintain a $\Sigma_{\text{surr}}$ that is proportional to their mass. This is especially true for the MMCs, which have masses that are proportional to their surrounding surface densities, regardless of clump evolutionary stage.  However in the same amount of time, other member cores will accrete less mass, while a steady stream of material from the intra-clump reservoir continues to increase their surrounding surface densities, which would reproduce the weaker $M_{\text{core}}$ vs. $\Sigma_{\text{surr}}$ correlation that we observe for clumps with higher $L/M$ ratios.

The trends seen in Figure \ref{fig:surfd_mcore} can also be interpreted through the lens of the fragmentation-induced starvation model, where the MMC initially in the center of a clump is shielded from additional mass accretion by cores further out that may have formed later in the clump's evolution \citep{Thomas_2010, Girichidis_2012}. Like the competitive accretion model, all cores gain mass throughout the evolution of the clump, but the rate at which the central MMC gains mass decreases as the accretion rate of the outer cores increases. In this scenario, the MMC masses and the number of lower mass cores located at higher local surface densities would still increase over time. It is difficult to distinguish the competitive accretion model from the starvation model, as it is unclear from Figure \ref{fig:surfd_mcore} how the mass accretion rates change for specific cores as time progresses.

Regardless, the trend observed in Figure \ref{fig:surfd_mcore} implies a dynamical, clump-fed accretion scenario, in which high-mass cores form from lower-mass ``seed" cores as the clump evolves. Our results are consistent with recent studies of core mass evolution across multiple surveys, such as ASSEMBLE \citep{Xu_2024}, ALMA-IMF \citep{2023_Pouteau}, and ASHES \citep{Morii_2024}. This conclusion also complements the recent findings from the ALMAGAL core catalog \citep{Coletta_2025}, as well as further analysis done on the spatial distribution of ALMAGAL cores (Schisano et al. 2025, submitted) and the relationship between ALMAGAL cores and their host clump (Elia et al.2025, submitted).

\subsection{Sources of Uncertainty}

We acknowledge that there are multiple sources of uncertainty in our measurements, in particular our mass estimations as they are highly reliant on our dust temperature assumptions. However, as discussed in Appendix \ref{sec:clump_temp_assumption}, if we instead assume the dust temperature of the structure is equal to the dust temperature of its host clump, $T_{\text{clump}}$, we find that the relationship shown in Figure \ref{fig:surfd_mcore} still holds, but exhibits a systematic shift towards higher surrounding surface densities. This makes sense as $T_{\text{clump}}$ is often less than the assumed core-scale temperature from the model we use in Section \ref{subsec:struc_prop}, which would result in higher masses, and consequently, surface densities for all structures. 

Ultimately, a thorough spectral line analysis of the cores and surrounding medium will be required to confirm the temperature and kinematics of each part of these systems. Future work, such as the core temperatures for the ALMAGAL sample (Jones et al. in prep.), will be essential for providing these measurements so that we can rigorously test a clump-fed theory of core mass accretion on our large sample of observations.

It is also possible that our results were impacted by our choice of dendrogram parameters. However, after generating several other versions of the catalog, we did not find any significant changes in our results. We provide more information on the range of parameters investigated and one explanatory figure in Appendix \ref{sec:dendro_param_error}.

We also explored how missing flux caused by short-spacings could impact our results. In Appendix \ref{sec:feathering}, we discuss single-dish combination using continuum observations from the ATLASGAL Survey \citep{Schuller_2009}, and quantitatively determine that for most structures, mass calculations differ by less than 10\% when using the single-dish combined data instead. Additionally, since feathered data includes more extended emission, our dendrogram structures would have higher measured surface densities while the core masses would remain largely the same. In this scenario, the range of local surface densities calculated from dendrogram structures would increase, causing a larger spread along the $x$-axis of Figure \ref{fig:surfd_mcore}, which is still consistent with the results we present in Section \ref{subsec:core_mass_accretion}. Given this information, and the fact that only 856 out of 900 targets have overlap with the ATLASGAL survey, we ultimately decided against using single-dish combined images in our analysis.

We also consider how the mass estimates of cores from the ALMAGAL core catalog may ultimately affect the results we present in Section \ref{subsec:core_mass_accretion}. We note that the presence of lower mass cores in more evolved regions may in part be due to the temperature model used (see Section \ref{subsec:struc_prop}), resulting in the underestimation of mass for more recently formed cores that are at a lower temperature than would be assumed based on the evolutionary stage of the clump. However, this same temperature model would likely also underestimate the cores surrounding surface density, so we consider the presence of lower mass cores at higher $\Sigma_{\text{surr}}$ to be a robust trend despite our temperature assumptions. Although the near sample targets have more low mass core detections than the far sample due to the difference in mass sensitivity, both groups sample a comparable number of clumps in the different evolutionary bins (see Table \ref{tab:sample}), so we consider this to have a negligible impact on the results shown in Figure \ref{fig:surfd_mcore}.

We also note that due to projection, some information on the 3D structure in our observations is lost.  For example, our clump mass estimate is based on an assumed symmetrical geometry for the clump in 3D, which is not necessarily the case across our sample. Similarly, with respect to our $\Sigma_{\text{surr}}$ calculations, it is possible for a core to be located outside of the central clump, but appear to be contained within a dense envelope in 2D projection. This would make it appear as though lower mass cores are contained in denser material, when in reality they are not. However, given the statistical significance of our sample we consider it unlikely that these projection effects would ultimately change the results of our analysis for either of the aforementioned scenarios.

\section{Conclusion}
\label{sec:conclusion}
In this paper, we report on the fundamental properties of the hierarchical continuum structures found in the ALMAGAL survey, and investigate how the extent of multi-scale fragmentation in 900 clumps relates to their evolutionary stage. We also analyze the connection between core mass as it relates to the surface density of its immediate environment and examine this as a function of clump evolution. 

The main findings of our analysis are summarized as follows:
\begin{enumerate}
    \item We produce common linear resolution (2000 au) continuum images for 904 out of 1013 continuum images with native resolutions $\leq 2000$ au. With these we are able to directly compare measurements made for continuum structures across the entire sample. 
    \item Using \verb|astrodendro|, we identify 5160 structures across 904 dense clumps targeted in the ALMAGAL survey. These structures have effective radii spanning $800 - 42000$ au and cover multiple orders of magnitude in mass, from $0.05 - 670$ M$_{\odot}$. 
    \item We utilize the dendrogram calculated for each clump to classify it according to its relative level of morphological complexity. Of the 900 clumps for which we have measured properties, we have 191 (21.2\%) Empty clumps, 445 (49.4\%) Isolated clumps, 187 (20.8\%) Simple clumps, and 77 (8.6\%) Rich clumps, in the order of increasing hierarchical fragmentation. 
    \item We find that clumps exhibiting a rich morphology have higher median dust temperatures, mass surface densities, MMC masses, and \textit{L/M} ratios -- all of which indicate that they represent a population of more evolved star-forming regions. 
    \item We investigate the connection between the MMC mass and its surrounding surface density and find that they are directly related for all clumps, regardless of  evolutionary stage, with more evolved clumps having a more extended range of possible MMC masses and local surface densities.
    \item We analyze the relationship between mass and surrounding surface density for all cores contained in the ALMAGAL core catalog, and compare it to the relationship using just the MMCs. Although there is still a positive correlation between these quantities, we find that non-MMCs can have proportionally higher surrounding surface densities. This behavior is exaggerated in evolved clumps, where lower mass cores can be located in higher local surface density regions than equivalent mass cores in clumps at earlier stages in their evolution. This could imply that cores are forming through a dynamical, clump-fed accretion process. 
\end{enumerate}

In this paper, we described the process for generating the ALMAGAL hierachical continuum structure catalog and presented the distribution of the fundamental properties of these structures. We found that the level of fragmented sub-structure in clumps is related to their evolution. Additionally, we used the hierarchical continuum structure to infer the mass evolution of cores in clumps at different evolutionary stages, finding evidence of a clump-fed accretion scenario. In future work, this hierarchical structure catalog will be a useful tool for assessing how properties such as the gas temperature and the abundance of molecular species changes across different spatial scales, providing context for the dynamical evolution of these star forming systems.

\section{Acknowledgments}

J.Wallace gratefully acknowledges funding from National Science Foundation under Award Nos. 2108938 and 2206510.
\\
C.\ Battersby  gratefully  acknowledges  funding  from  National  Science  Foundation  under  Award  Nos. 2108938, 2206510, and CAREER 2145689, as well as from the National Aeronautics and Space Administration through the Astrophysics Data Analysis Program under Award ``3-D MC: Mapping Circumnuclear Molecular Clouds from X-ray to Radio,” Grant No. 80NSSC22K1125.
\\
A.S.-M.\ acknowledges support from the RyC2021-032892-I grant funded by MCIN/AEI/10.13039/501100011033 and by the European Union `Next GenerationEU’/PRTR, as well as the program Unidad de Excelencia María de Maeztu CEX2020-001058-M, and support from the PID2023-146675NB-I00 (MCI-AEI-FEDER, UE).
\\
G.A.F gratefully acknowledges the Deutsche Forschungsgemeinschaft  (DFG) for funding through SFB 1601 ``Habitats of massive stars across cosmic time'' (sub-project B1)  and from the University of Cologne and its Global Faculty programme. 
\\
L.B.  gratefully acknowledges support by the  ANID BASAL project FB210003
\\
P.S. was partially supported by a Grant-in-Aid for Scientific Research (KAKENHI Number JP23H01221) of JSPS.
\\
R.K. acknowledges financial support via the Heisenberg Research Grant funded by the Deutsche Forschungsgemeinschaft (DFG, German Research Foundation) under grant no.~KU 2849/9, project no.~445783058.
\\
T. Z.\ acknowledges funding from NSFC (Grant No. 12373026), the Leading Innovation and Entrepreneurship Team of Zhejiang Province of China (Grant No. 2023R01008), the Key R\&D Program of Zhejiang, China (Grant No. 2024SSYS0012).
\\
The Teams at INAF-IAPS acknowledge financial support from the European Research Council via the ERC Synergy Grant ``ECOGAL'' (project ID 855130).
\\
R.S.K.\ acknowledges financial support from the ERC via Synergy Grant ``ECOGAL'' (project ID 855130),  from the German Excellence Strategy via the Heidelberg Cluster ``STRUCTURES'' (EXC 2181 - 390900948), and from the German Ministry for Economic Affairs and Climate Action in project ``MAINN'' (funding ID 50OO2206).  
\\
This paper makes use of the following ALMA data: ADS/JAO.ALMA\#2019.1.00195.L. ALMA is a partnership of ESO (representing its member states), NSF (USA) and NINS (Japan), together with NRC (Canada), MOST and ASIAA (Taiwan), and KASI (Republic of Korea), in cooperation with the Republic of Chile. The Joint ALMA Observatory is operated by ESO, auI/NRAO and NAOJ.
\\
This research made use of astrodendro, a Python package to compute dendrograms of Astronomical data ~(\url{http://www.dendrograms.org/}).
\\
Part of this research was carried out at the Jet Propulsion Laboratory, California Institute of Technology, under a contract with the National Aeronautics and Space Administration (80NM0018D0004).


\vspace{5mm}
\facilities{ALMA, APEX}

\software{Numpy \citep{harris2020array}, Scipy \citep{2020SciPy-NMeth}, Astropy \citep{astropy_2013, astropy_2018, astropy_2022}, Astrodendro (\url{http://www.dendrograms.org/}), Pandas \citep{reback2020pandas}, CASA: version 6.5.5.21 \citep{The_CASA_Team_2022}
           }



\appendix

\section{Clump Statistics For Each Morphological Category}
\label{clump_stat}

In Table \ref{clump_stat_tab}, we summarize the statistical quantities for the violin plot distributions shown in Section \ref{subsec:clump_distributions}.

\begin{deluxetable}{cccccccccccccccc}[h]
\tabletypesize{\scriptsize}
\tablecaption{A summary of the clump statistics for the different sub-samples (Rich, Isolated, Simple, Rich and Empty). The 25th (Q1), 50th (Q2), and 75th (Q3) percentile values for each distribution is reported. \label{clump_stat_tab}}
\startdata
\tablehead{ \colhead{} & \colhead{} & \colhead{All} & \colhead{} & \colhead{} & \colhead{Empty} &\colhead{}  & \colhead{} & \colhead{Isolated} & \colhead{} & \colhead{} & \colhead{Simple} & \colhead{}&  \colhead{}& \colhead{Rich} & \colhead{ }}
Property & Q1 & Q2 & Q3 & Q1 & Q2 & Q3 & Q1 &Q2 &Q3 & Q1 & Q2 & Q3 & Q1 & Q2 & Q3 \\
\hline
$d$ (pc) & 3480.0 & 4570.0 & 5500.0 & 3955.0 & 5160.0 & 5695.0 & 3390.0 & 4400.0 & 5350.0 & 3320.0 & 4120.0 & 5310.0 & 3890.0 & 4870.0 & 5790.0 \\
$R_{\text{GCR}}$ (pc) & 4377.4 & 5115.2 & 6377.4 & 4436.5 & 5931.6 & 8376.9 & 4396.6 & 5123.2 & 6319.8 & 4401.1 & 5088.8 & 6161.4 & 4217.2 & 4837.9 & 5949.6 \\
$M_{\text{clump}}$ ($M_{\odot}$) & 570.9 & 779.7 & 1190.9 & 544.6 & 711.7 & 1085.2 & 556.5 & 736.3 & 1051.6 & 623.4 & 810.7 & 1275.3 & 892.5 & 1308.1 & 1782.1 \\
$L/M$ ($L_{\odot}/M_{\odot}$)& 0.3 & 1.5 & 10.4 & 0.2 & 0.7 & 1.9 & 0.2 & 0.9 & 6.7 & 1.3 & 5.6 & 17.8 & 8.5 & 20.7 & 41.4 \\
$\Sigma_{\text{clump}}$ (g cm$^{-2}$) & 0.4 & 0.6 & 1.1 & 0.2 & 0.3 & 0.6 & 0.4 & 0.6 & 0.9 & 0.6 & 0.9 & 1.4 & 0.8 & 1.3 & 1.8 \\
$M_{\text{MMC}}$ ($M_{\odot}$)& 0.3 & 1.0 & 3.0 & 0.0 & 0.0 & 0.0 & 0.5 & 0.9 & 2.1 & 1.5 & 3.0 & 5.4 & 3.2 & 5.4 & 9.3 \\
$T_{\text{clump}}$ (K) & 11.4 & 14.3 & 19.7 & 10.7 & 12.2 & 14.5 & 11.0 & 13.2 & 17.6 & 14.4 & 18.7 & 22.4 & 21.3 & 24.0 & 28.4 \\
\enddata
\end{deluxetable}

\section{Effect of dendrogram parameters on results}
\label{sec:dendro_param_error}

\begin{table}
\begin{center}
\caption{A summary of the dendrogram parameters used for different versions of the catalog.}
\begin{tabular}{cccc}
\hline\hline
$f_{\text{min}}$ & $\delta$ & $f_{\text{peak}}$ & $n_{\text{pix}}$ \\
\hline
2$\sigma_{\text{rms}}$ & 3$\sigma_{\text{rms}}$ & 3$\sigma_{\text{rms}}$ & 0.75$\times A_{\text{beam,pix}}$ \\
2$\sigma_{\text{rms}}$ & 3$\sigma_{\text{rms}}$ & 3$\sigma_{\text{rms}}$ & 1.0$\times A_{\text{beam,pix}}$  \\
2$\sigma_{\text{rms}}$ & 3$\sigma_{\text{rms}}$ & 2$\sigma_{\text{rms}}$ & 0.5$\times A_{\text{beam,pix}}$  \\
2$\sigma_{\text{rms}}$ & 3$\sigma_{\text{rms}}$ & 4$\sigma_{\text{rms}}$ & 0.5$\times A_{\text{beam,pix}}$ \\
2$\sigma_{\text{rms}}$ & 2$\sigma_{\text{rms}}$ & 3$\sigma_{\text{rms}}$ & 0.5$\times A_{\text{beam,pix}}$  \\
2$\sigma_{\text{rms}}$ & 4$\sigma_{\text{rms}}$ & 3$\sigma_{\text{rms}}$ & 0.5$\times A_{\text{beam,pix}}$  \\
3$\sigma_{\text{rms}}$ & 3$\sigma_{\text{rms}}$ & 3$\sigma_{\text{rms}}$ & 0.5$\times A_{\text{beam,pix}}$  \\
\hline
\end{tabular}
\label{tab:dendro_versions}
\end{center}
\end{table}

\begin{figure}[htb!] 
\begin{centering}
\epsscale{1.0}
\plotone{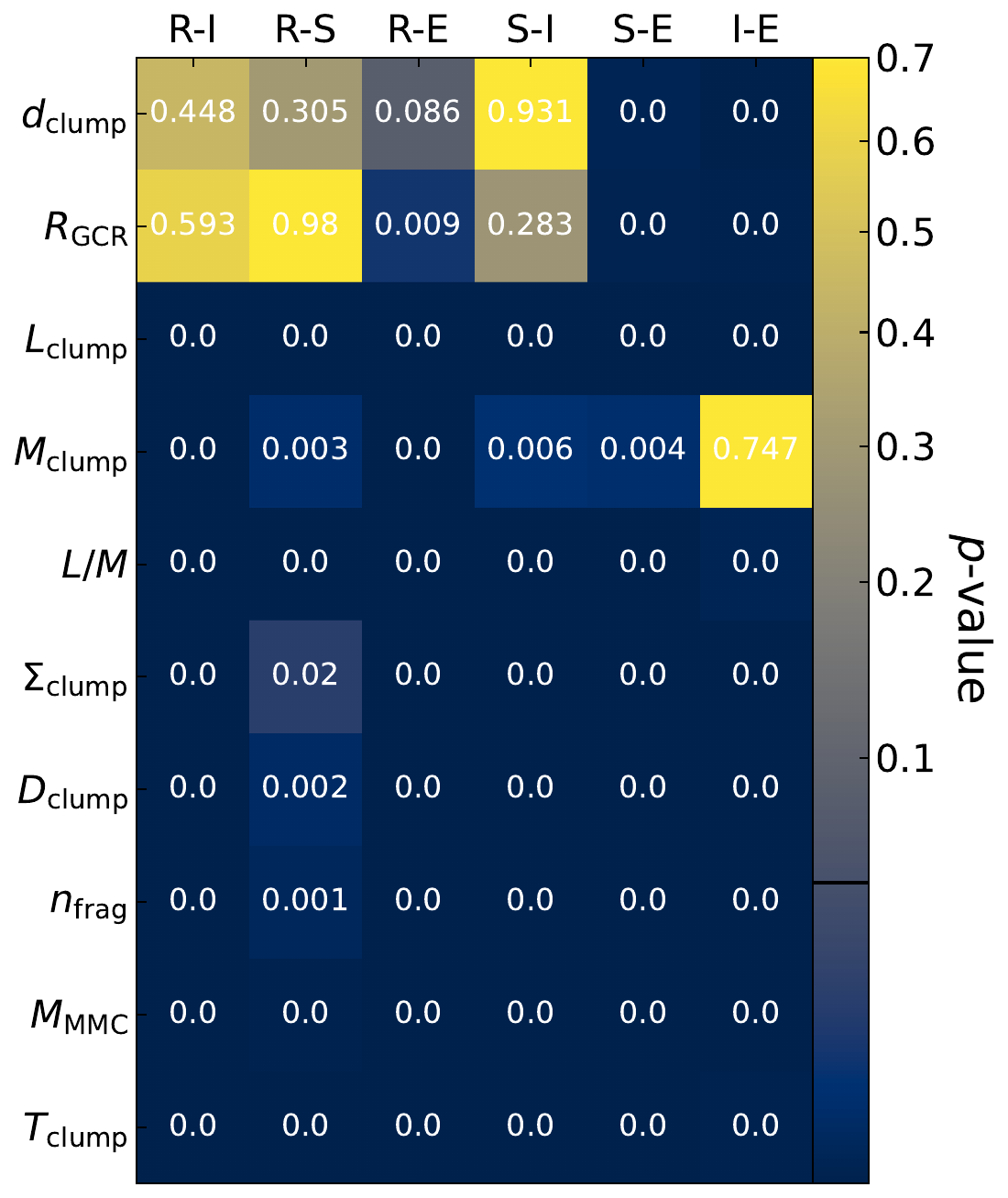}
\caption{A recreation of Figure \ref{fig:ks_tests}  Heatmap matrix made using a version of the hierarchical structure catalog made using a minimum value of $f_{\text{min}}$ of 3$\sigma_{\text{rms}}$ instead of 2$\sigma_{\text{rms}}$. All \textit{p}-values are shown in white text. A \textit{p}-value $= 0.05$ is shown with a solid black line on the color bar. The \textit{p}-values $< 0.05$ indicate combinations where we reject the null hypothesis that both samples are pulled from an identical distribution at a confidence level of 95\%. The clump properties are listed on the $y$-axis, where $d_{\text{clump}}$ is the heliocentric distance, $R_{\text{GCR}}$ is the Galactocentric radius, $L_{\text{clump}}$ is the clump luminosity, $M_{\text{clump}}$ is the clump mass, $L/M$ is the clump luminosity-to-mass ratio, $\Sigma_{\text{clump}}$ is the clump surface density, $D_{\text{clump}}$ is the clump diameter, $n_{\text{frag}}$ is the fragment density of cores from the ALMAGAL core catalog \citep{Coletta_2025} in the clump (the number of cores divided by the area of the observed field of view), and $T_{\text{clump}}$ is the clump dust temperature.}
\label{fig:ks_tests_dendro_test}
\end{centering}
\end{figure}

\begin{figure*}[!h] 
\epsscale{1}
\begin{centering}
\plottwo{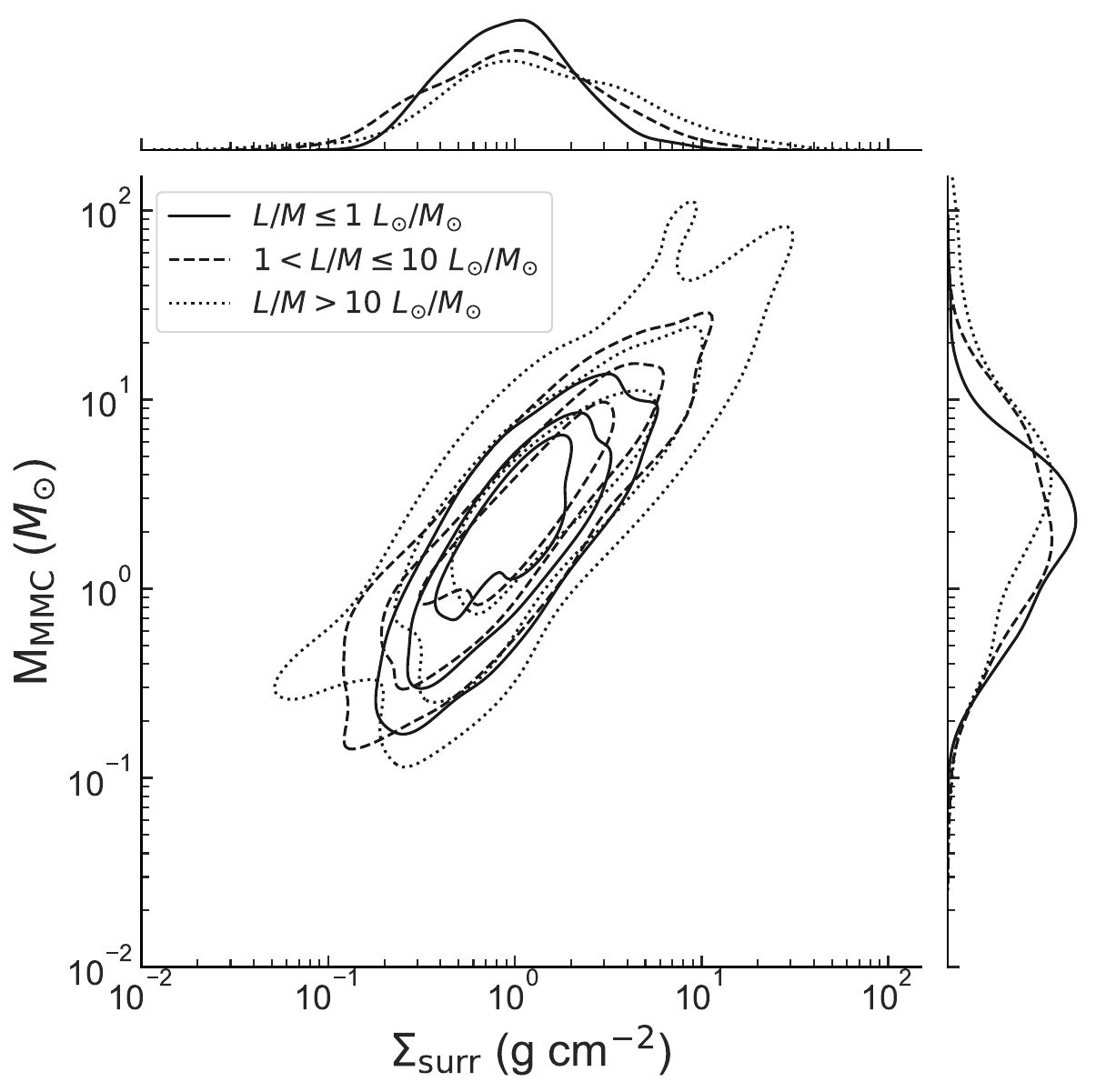}{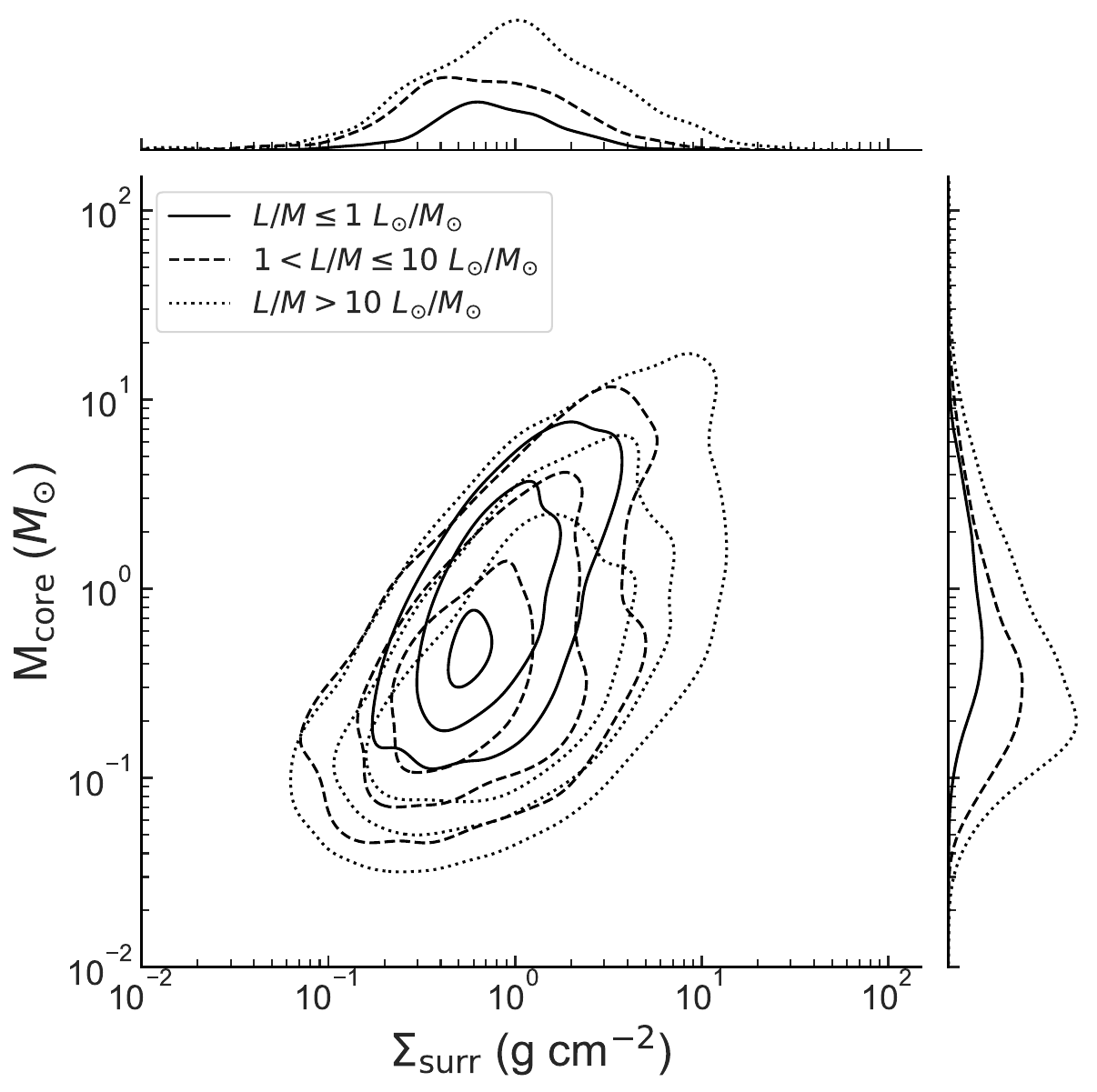}
\caption{A recreation of Figure \ref{fig:surfd_mcore} made using a version of the hierarchical structure catalog made using a minimum value of $f_{\text{min}}$ of 3$\sigma_{\text{rms}}$ instead of 2$\sigma_{\text{rms}}$. As in Figure \ref{fig:surfd_mcore}, the contours indicate the shape of the distribution for the inner [90, 70, 40]\% of the data.  Marginal plots indicate the 1D kernel density estimates of the dendrogram structure surface density and core mass on the \textit{x}- and \textit{y}-axes, respectively.}
\label{fig:dendro_param_test}
\end{centering}
\end{figure*}

To determine how the results are affected by our choice of dendrogram parameters, we generated several new versions of the catalog. We list the parameters used for each version in Table \ref{tab:dendro_versions}. All versions of the catalog do not change the main results of our paper. In Figures \ref{fig:ks_tests_dendro_test} and \ref{fig:dendro_param_test}, we present the KS test results and the $\Sigma_{\text{surr}}-M_{\text{core}}$ relationship for the version of the catalog that had the most significant impact on our results. This version has the following parameters: $f_{\text{min}} = 3\sigma_{\text{rms}}, \delta=3\sigma_{\text{rms}}, f_{\text{peak}} = 3\sigma_{\text{rms}}, n_{\text{pix}}=0.5 \times A_{\text{beam,pix}}$. The primary difference between Figures \ref{fig:ks_tests} and \ref{fig:ks_tests_dendro_test} is that the $p$-values for $d_{\text{clump}}$ and $R_{\text{GCR}}$ increase overall (excluding pairs that include the Empty subsample.) In Figure \ref{fig:dendro_param_test}, we find no significant change in the relationship between $\Sigma_{\text{surr}}$ and $M_{\text{core}}$.

\section{The $\Sigma_{\text{surr}} -M_{\text{core}}$ relation using clump temperature assumption} 
\label{sec:clump_temp_assumption}

\begin{figure*}[!h] 
\epsscale{1}
\begin{centering}
\plottwo{ALMAGAL_masssigma_mmcmass_update_2_ms3_mp0.5_pt3.pdf}{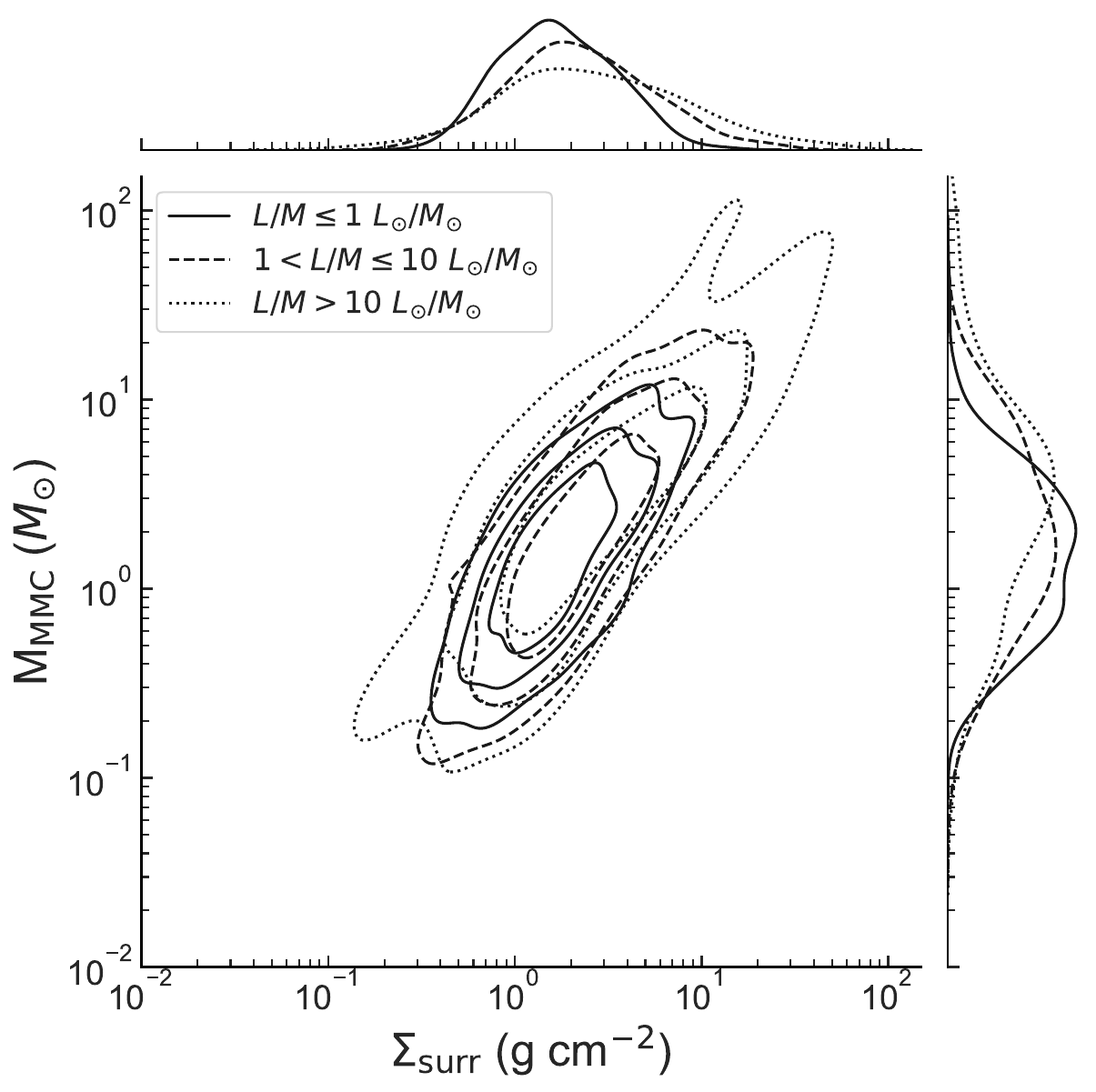}
\plottwo{ALMAGAL_masssigma_mcore_evstages_update_2_ms3_mp0.5_pt3.pdf}{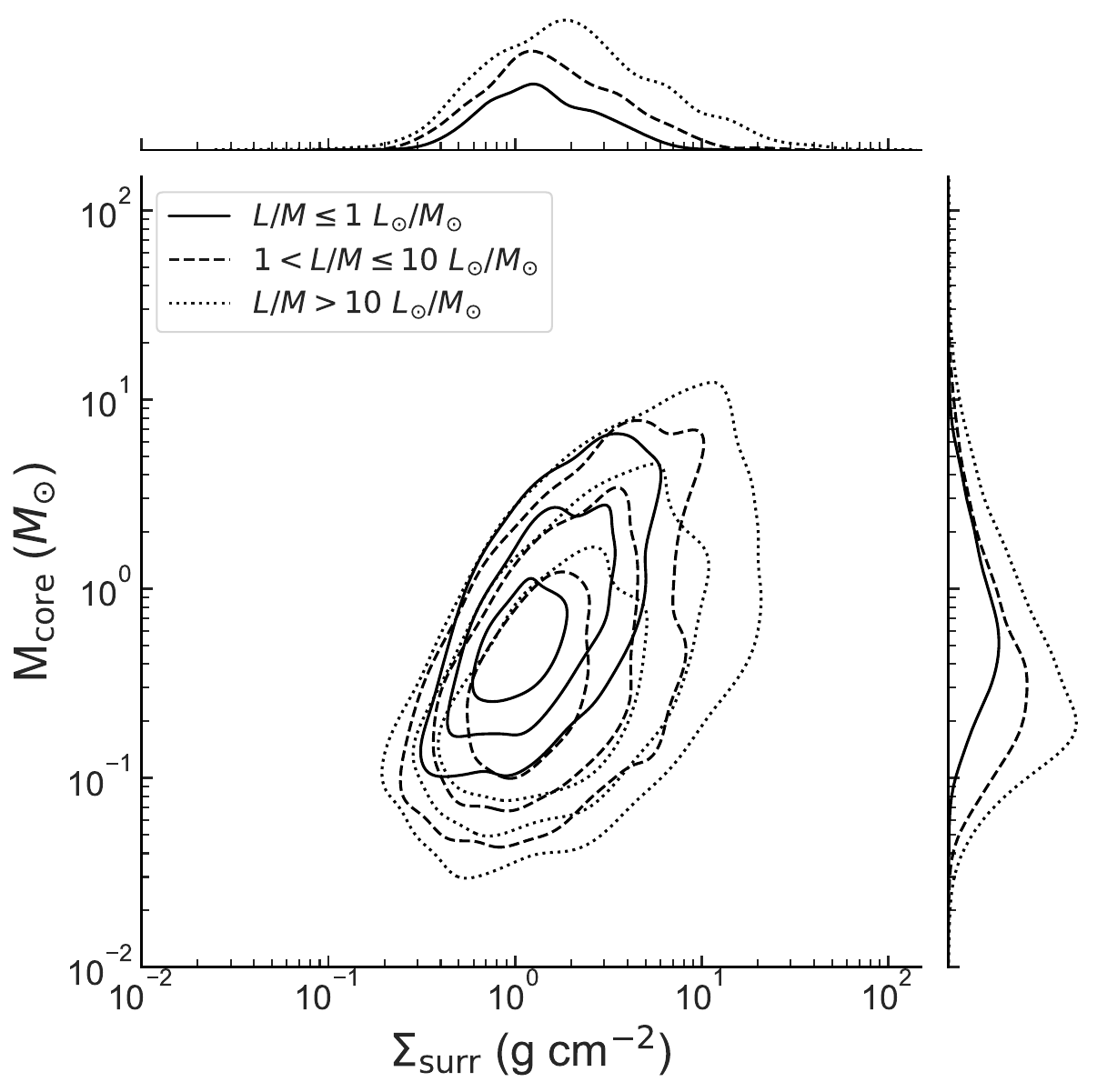}
\caption{A comparison of the results in Figure \ref{fig:surfd_mcore} when using $\Sigma_{\text{surr}}$ values calculated using the temperature model presented in Section \ref{subsec:struc_prop} the non-feathered images (left panels) and the feathered images (right panels). As in Figure \ref{fig:surfd_mcore}, the contours indicate the shape of the distribution for the inner [90, 70, 40]\% of the data.  Marginal plots indicate the 1D kernel density estimates of the dendrogram structure surface density and core mass on the \textit{x}- and \textit{y}-axes, respectively.}
\label{fig:surfd_mcore_temp_comp}
\end{centering}
\end{figure*}

For the results we present in Section \ref{subsec:core_mass_accretion} we use the temperature model presented in Section \ref{subsec:struc_prop} to calculate $\Sigma_{\text{surr}}$ for dendrogram structures surrounding cores from the ALMAGAL core catalog. To better understand how our temperature assumption affects our results, we decided to make this calculation using the clump-scale temperature, $T_{\text{clump}}$, as well. In Figure \ref{fig:surfd_mcore_temp_comp}, we present the $\Sigma_{\text{surr}} -M_{\text{core}}$ relation using our original temperature model as well as the clump temperature assumption. In general, we find that $\Sigma_{\text{surr}}$ values increase systematically, which makes sense since the lower assumed temperatures from the clump-scale will result in higher mass estimates for the structures. In general, we find that the general trends still hold, and that the extents observed in $M_{\text{core}}$ vs. $\Sigma_{\text{surr}}$ still depend on the evolutionary stage of the hosting clump.

\newpage
\section{Mass Calculations Using Feathered Data}
\label{sec:feathering}

Millimeter interferometry allows astronomers to make observations at very high, sub-arcsecond resolution. However, these observations filter out larger scale emission due to the missing short-baseline data. As a solution to this, one can combine interferometric data with single dish data to produce an image that is sensitive to both large and small scales and retains the true flux along the line-of-sight. 

We combined the interferometric ALMAGAL continuum images with 870 $\micron$ ATLASGAL continuum observations with an angular resolution of 19\arcsec. To do this, we implemented the CASA \verb|feather| task. In our ALMAGAL target sub-sample (900 targets), there are 856 observations that overlap with the ATLASGAL observation footprint.

These ALMAGAL and ATLASGAL continuum observations are centered at different wavelengths, so we apply a correction factor to the ATLASGAL images assuming that the emission can be modelled with a blackbody modified by the dust opacity $k_{\nu}\propto \beta$:

\begin{equation} \label{scale_factor}
    f = \frac{F_{\text{ALMA}}}{F_{\text{ATLASGAL}}} = \left( \frac{\nu_{\text{ALMA}}}{\nu_{\text{ATLASGAL}}} \right)^{\alpha_{\nu}} = \left( \frac{219}{345} \right)^{3.5} \approx 0.2
\end{equation}

The ALMAGAL and ATLASGAL data are centered at frequencies of $\sim$ 219 GHz and $\sim$ 345 GHz, respectively. If we use Equation \ref{scale_factor} (with $\beta = 1.5, \alpha_{\nu} = 3.5$), we acquire a scale factor of $f \sim$ 0.2, which is fairly small. 

We directly compare the mass calculations for each dendrogram structure using the feathered images and the original interferometric images. We also consider how this metric changes when using the background subtracted mass values. For each structure, we calculate a percent mass difference: 

\begin{equation}
    \frac{|M_{\text{feather}}-M_{\text{ALMA}}|}{M_{\text{feather}}} \times 100.
\end{equation}

\begin{figure}
\begin{centering}
\plottwo{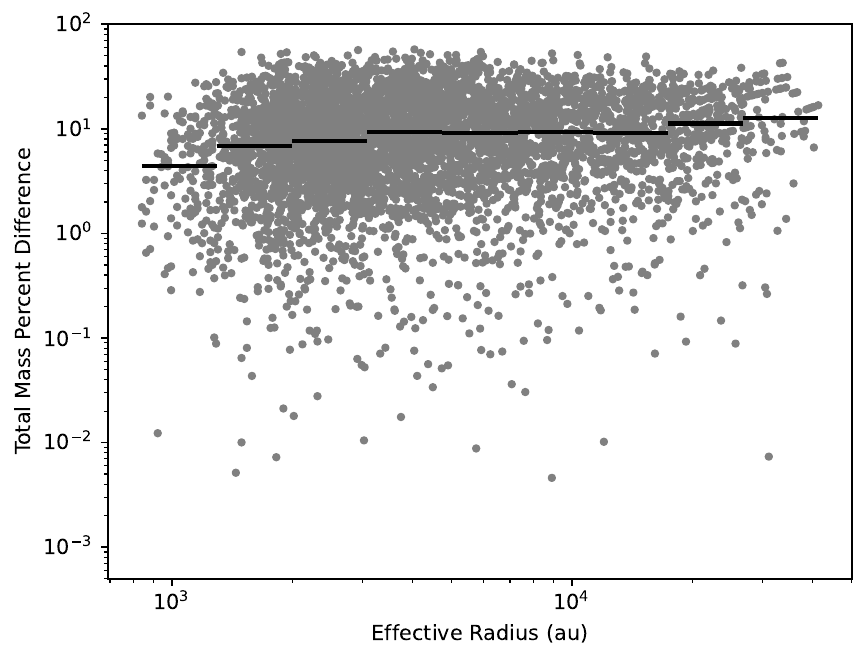}{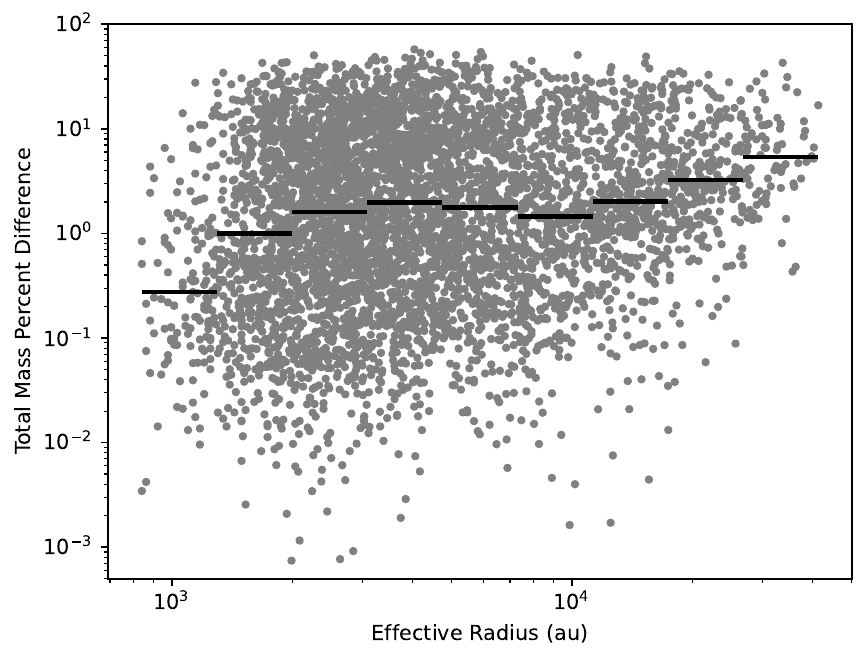}
\caption{The total percent difference when comparing structure masses using the feathered images and the original interferometric ALMA images as a function of the effective radius of the structure in au. The left panel uses the non-background subtracted masses and the right panel uses the background subtracted masses (Note: We do not calculate a background-subtracted mass for trunk (level=0) structures.) The horizontal black lines indicate the median mass percent value for each radius bin.} 
\label{fig:feather_mass_diff}
\end{centering}
\end{figure}

In Figure \ref{fig:feather_mass_diff}, we examine the percent mass difference as a function of $R_{\text{eff}}$ for each structure. For the non-background subtracted masses, we see that the median mass percent difference is between $4-12\%$ across all $R_{\text{eff}}$ bins, with a slight increase with increasing size. The majority ($\sim72\%$) of structures have a mass percent difference $<15\%$.  For the background subtracted masses, the mass percent difference is between $0.3-5\%$, and $\sim 89\%$ of structures have a mass percent difference $<15\%$.

In light of this information, we have decided to use the 7M+TM1+TM2 ALMA continuum data without single dish combination. One reason is that the feathered images are generated from primary beam corrected data. Our source extraction procedure works best with non-primary beam corrected images, since the noise does not change across the image. If we run dendrogram source extraction on the feathered data, we will be extracting a large quantity of spurious detections at the edge of the map that will be difficult to remove from the catalog. Additionally, we would also be working with a reduced sample size, as 44 ALMAGAL targets have no overlap with the ATLASGAL images. Finally, considering that the feathered outputs have a minimal impact on the masses of structures in our catalog, we determine that using the original ALMAGAL images without single dish combination is the best choice for the analysis put forth in this paper.  

\begin{figure*}[!htb] 
\epsscale{1}
\begin{centering}
\plottwo{ALMAGAL_masssigma_mmcmass_update_2_ms3_mp0.5_pt3.pdf}{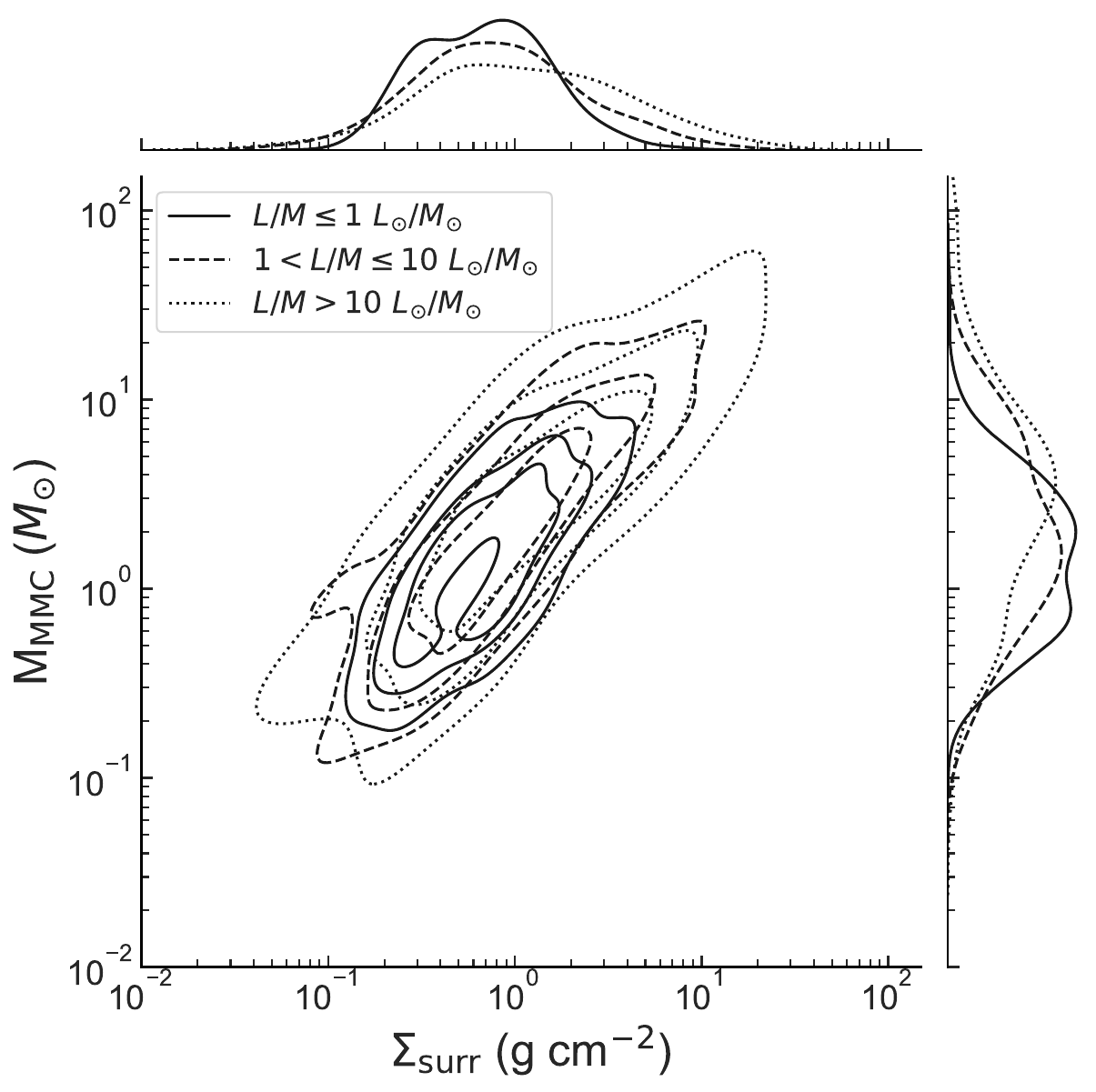}
\plottwo{ALMAGAL_masssigma_mcore_evstages_update_2_ms3_mp0.5_pt3.pdf}{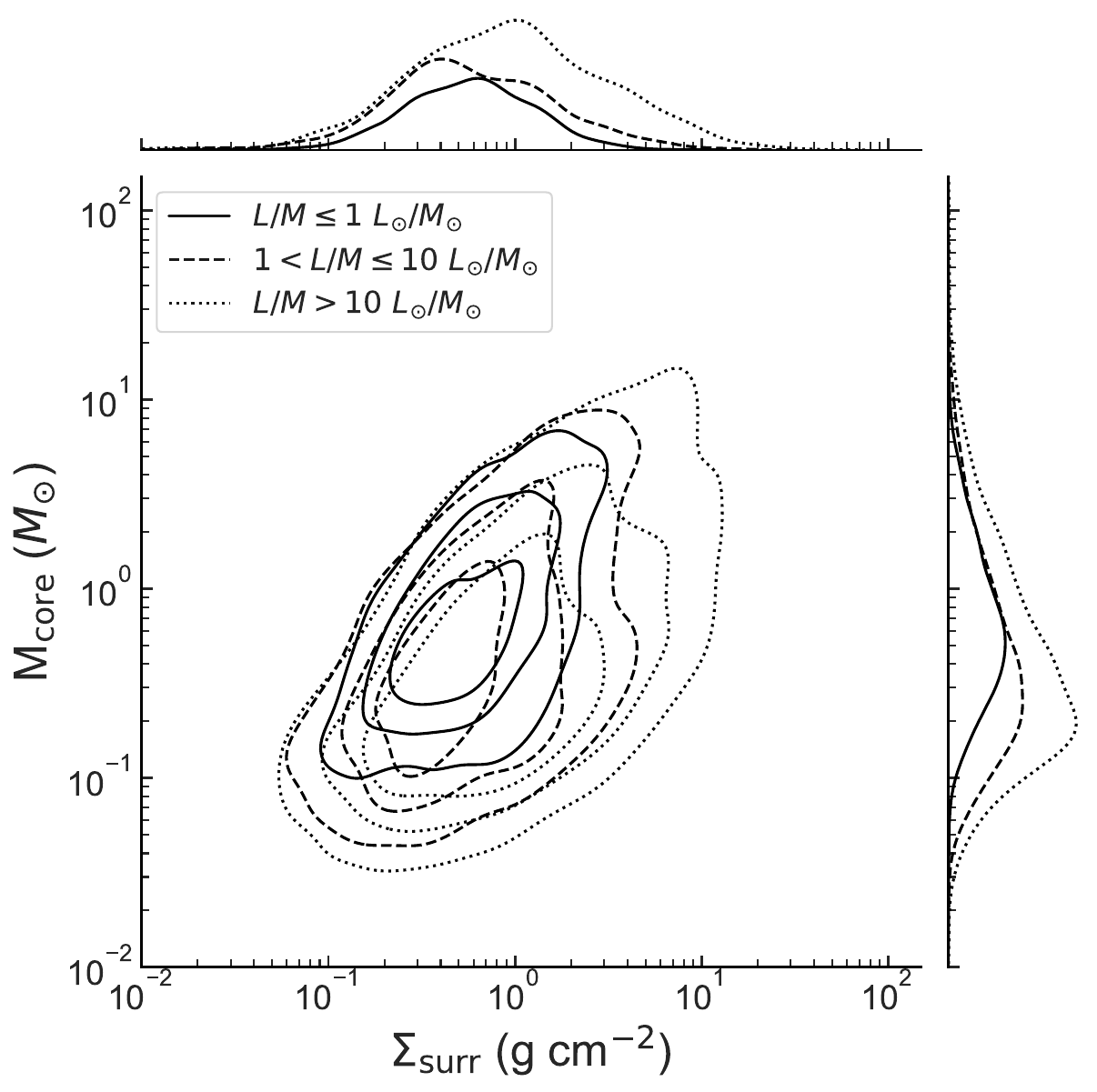}
\caption{A comparison of the results in Figure \ref{fig:surfd_mcore} when using the non-feathered images (left panels) and the feathered images (right panels). As in Figure \ref{fig:surfd_mcore}, the contours indicate the shape of the distribution for the inner [90, 70, 40]\% of the data.  Marginal plots indicate the 1D kernel density estimates of the dendrogram structure surface density and core mass on the \textit{x}- and \textit{y}-axes, respectively.}
\label{fig:surfd_mcore_feather_comp}
\end{centering}
\end{figure*}

\newpage
\section{Structure differences in the near and far sample}
\label{sec:near_far_mvr_diff}

\begin{figure*}[htb!] 
\begin{centering}
\epsscale{1.1}
\plotone{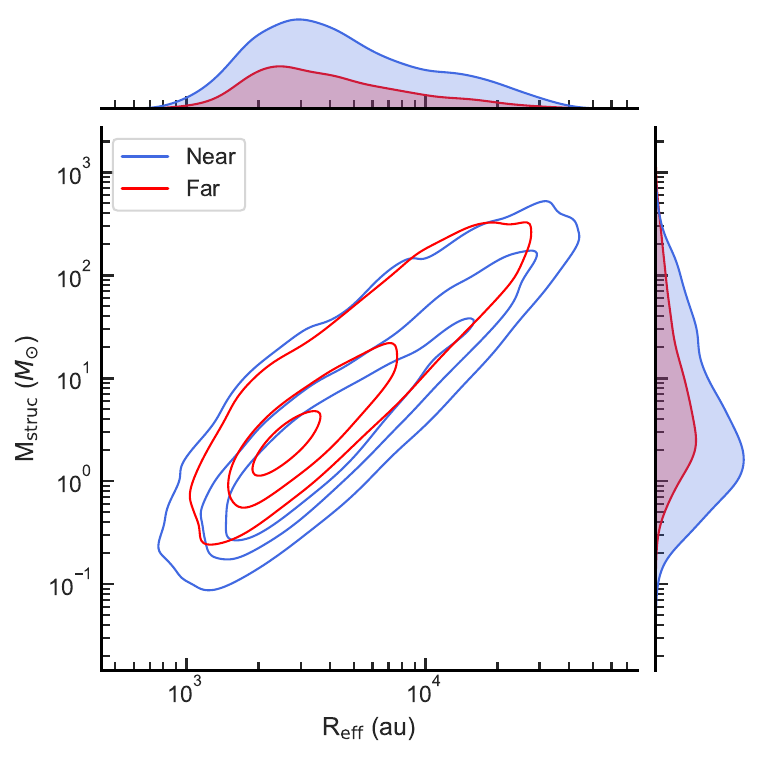}
\caption{A 2D kernel density estimate comparing the mass-radius relation for near (blue) and far samples (red) structures in the ALMAGAL hierarchical structure catalog. The contours indicate the shape of the distributions for the inner [90, 50, 20]\% of the data.}
\label{fig:mvr_nearfar}
\end{centering}
\end{figure*}

As discussed in Section \ref{subsec:clump_class}, the near and far sample distributions exhibit a systematic difference in mass sensitivity, with the far sample being less sensitive to lower mass structures. Since the dendrogram output is dependent on the mass sensitivity (see Section \ref{subsec:clump_class}), we ultimately  extract less low-mass structures from the far sample images, as seen in Figure \ref{fig:mvr_nearfar}. The difference in mass sensitivity can also explain why the near sample contains the largest structures, despite the fact that the linear size for the near sample field of view is generally smaller than that of the far sample.

\bibliography{ref}{}
\bibliographystyle{aasjournal}



\end{document}